\documentclass[journal]{IEEEtran}
\usepackage{graphicx}
\usepackage{subfigure}
\usepackage{array}
\usepackage{amsmath}
\usepackage{amssymb}
\usepackage{multirow}
\usepackage{cases}
\usepackage{cite}
\usepackage{fancyhdr}
\usepackage{color}
\usepackage[justification=centering]{caption}
\usepackage[colorlinks=true,citecolor=black,linkcolor=black,anchorcolor=black,urlcolor=black,bookmarks=false]{hyperref}
\UseRawInputEncoding
\ifCLASSINFOpdf
\else
\fi
\begin{document}
\title{No-Reference Light Field Image Quality Assessment Based on Spatial-Angular Measurement}

\author{Likun Shi, Wei Zhou,~\IEEEmembership{Student Member,~IEEE}, Zhibo Chen,~\IEEEmembership{Senior Member,~IEEE}, and Jinglin Zhang
        %,~\IEEEmembership{Fellow,~OSA,}
       % and~Jane~Doe,~\IEEEmembership{Life~Fellow,~IEEE}% <-this % stops a space
\thanks{L. Shi, W. Zhou and Z. Chen are with the CAS Key Laboratory of Technology in Geo-Spatial Information Processing and Application System, University of Science and Technology of China, Hefei 230027, China, (e-mail: slikun@mail.ustc.edu.cn; weichou@mail.ustc.edu.cn; chenzhibo@ustc.edu.cn).}
\thanks{J. Zhang is with Key Laboratory of Meteorological Disaster, Ministry of Education, Nanjing University of Information Science and Technology, Nanjing 210044, China, (e-mail: jinglin.zhang@nuist.edu.cn).}
\thanks{Likun Shi and Wei Zhou contributed equally to this paper. Corresponding Authors: Zhibo Chen, Jinglin Zhang.}
\thanks{This work was supported in part by NSFC under Grant 61571413, 61632001.}
% <-this % stops a space
}



\markboth{IEEE Transactions on Circuits and Systems for Video Technology}%
{Shell \MakeLowercase{\textit{et al.}}: Bare Demo of IEEEtran.cls for IEEE Journals}




\maketitle

\begin{abstract}
Light field image quality assessment (LFI-QA) is a significant and challenging research problem. It helps to better guide light field acquisition, processing and applications.
However, only a few objective models have been proposed and none of them completely consider intrinsic factors affecting the LFI quality.
In this paper, we propose a No-Reference Light Field image Quality Assessment (NR-LFQA) scheme, where the main idea is to quantify the LFI quality degradation through evaluating the spatial quality and angular consistency.
We first measure the spatial quality deterioration by capturing the naturalness distribution of the light field cyclopean image array, which is formed when human observes the LFI.
Then, as a transformed representation of LFI, the Epipolar Plane Image (EPI) contains the slopes of lines and involves the angular information. Therefore, EPI is utilized to extract the global and local features from LFI to measure angular consistency degradation.
Specifically, the distribution of gradient direction map of EPI is proposed to measure the global angular consistency distortion in the LFI. We further propose the weighted local binary pattern to capture the characteristics of local angular consistency degradation.
Extensive experimental results on four publicly available LFI quality datasets demonstrate that the proposed method outperforms state-of-the-art 2D, 3D, multi-view, and LFI quality assessment algorithms.
\end{abstract}

\begin{IEEEkeywords}
Light field image, quality assessment, epipolar plane image, naturalness, spatial quality, angular consistency.
\end{IEEEkeywords}

% \fi
\IEEEpeerreviewmaketitle

\section{Introduction}
%\IEEEPARstart{T}{his} demo file is intended to serve as a ``starter file''
%for IEEE journal papers produced under \LaTeX\ using
%IEEEtran.cls version 1.8b and later.
%I wish you the best of success.

\IEEEPARstart{W}{ith} the rapid development of acquisition, transmission and display technologies, many new media modalities (e.g. virtual reality and light field) have been created to provide end-users with better immersive viewing experiences \cite{tech2016overview,mekuria2016design}. Unlike traditional media technologies, such as 2D image or stereoscopic image, light field content contains a large amount of information, which not only records radiation intensity information, but also records the direction information of light rays in the free space.
Due to abundance of spatial and angular information, LF research has attracted widespread attention and has many applications, such as rendering new views, depth estimation, refocusing and 3D modeling \cite{wu2017light}.

Distribution of light rays is first described by Gershun \textit{et al.} in 1939 \cite{gershun1939light}, and then Adelson and Bergen refined its definition and explicitly defined the model called plenoptic function in 1991, which describes the light rays as a function of intensity, location in space, travel direction, wavelength, and time \cite{adelson1991plenoptic}. However, obtaining and dealing with this multidimensional function is a huge challenge. To solve this problem and for practical usage, the measurement function is assumed to be monochromatic, time-invariant and the ray radiation remains constant along the line. The light field model can be simplified to a 4D function that can be parameterized as $L(u,v,s,t)$, where $(u,v)$ represents angular coordinate and $(s,t)$ expresses the spatial coordinate \cite{levoy1996light,gortler1996lumigraph}. Fig. 1 illustrates a light field image captured by Lytro Illum \cite{lytro_cam}, where each image in the LFI is called a Sub-Aperture Image (SAI). Additionally, we can produce EPI by fixing a spatial coordinate ($s$ or $t$) and an angular coordinate ($u$ or $v$). Specifically, vertical EPI can be obtained by fixing $u$ and $s$, similarly fixing $v$ and $t$ to get the horizontal EPI, as shown in the bottom and right of Fig. 1.

According to the above mentioned light field function, many light field capture equipments have been designed, and some of them have entered the market.
In general, light field acquisition approaches can be categorized into High Density Camera Array (HDCA), Time-Sequential Capture (TSC) and Micro-Lens Array (MLA) \cite{wu2017light}.
Specifically, the HDCA \cite{jpeg2017switer} establishes an array of image sensors distributed on a plane or sphere to simultaneously capture light field samples from different viewpoints \cite{levoy1996light}.
The angular dimensions are determined by the number of cameras and their distribution, while the spatial dimensions are determined by camera sensor.
The TSC system is designed that only uses a single image sensor to capture multiple samples of light field through multiple exposures, which captures light field at different viewpoints by moving the image sensor \cite{unger2003capturing} or fixing the camera when moving objects \cite{tamboli2016super}.
The angular dimensions are determined by the number of exposures, while the spatial dimensions are determined by camera sensor. In some cases (e.g. negligible illumination changes), the LFI captured by the TSC can be considered to be the same as that captured by the HDCA.
Different from the structure of the above devices, the MLA (also called lenslet array) system encodes 4D LFIs into a 2D sensor plane and has been successfully commercialized \cite{raytrix20173d,lytro_cam}.
%In the Lytro Illum camera \cite{lytro_cam}, the image sensor is placed at the focal length of the micro-lenes, and MLA is inserted aims to separate the rays of light based on direction, creating a focused image of the aperture of the main lens on the array of pixels underneath the micro-lens.
Fig. 2(a) exhibits the output of the Lytro Illum camera, which is called lenslet image and consists of many micro-images produced by the micro-lens (as shown in Fig. 2(b)). Moreover, the lenslet image can be further converted into multiple SAIs, as shown in Fig. 1.
The spatial resolution is defined by the number of micro-lens, and the number of pixels behind a micro-lens defines the angular resolution.

Benefiting from the development of the aforementioned acquisition equipment and the fast evaluation of 5G technologies, we can foresee the explosion of immersive media applications presented to end-customers. Specifically, light field display architectures are generally classified into three categories: traditional light field displays \cite{masia2013survey}, multilayer light field displays \cite{narain2015optimal} and compressive light field displays \cite{maimone2013focus}. Moreover, light field images can be visualized in two ways, namely integral light field structure (i.e. 2D array of images) and 2D slices, such as the EPI. With a sufficiently dense set of cameras, one can create virtual renderings of light field at any position of the sphere surface or even closer to the object by resampling and interpolating light rays \cite{levoy1996light}, rather than synthesizing the view based on geometry information \cite{mcmillan1995plenoptic}. The light field quality monitoring plays an important role in the processing and applications of the LFI for providing users with a good viewing experience. However, most LFI processing and applications do not consider specific characteristic of LFI or only utilize traditional 2D image quality assessment (IQA) metrics, which ignore the relationship between different SAIs.

Existing research indicates that LFI quality is affected by three factors, namely spatio-angular resolution, spatial quality, and angular consistency \cite{wu2017light,amirpour2019reliability}. The spatio-angular resolution refers to the number of SAIs in a LFI and the resolution of a SAI. Moreover, the spatial quality indicates SAI quality, while the angular consistency measures the visual coherence between SAIs. Since spatio-angular resolution is determined by acquisition devices, this paper focuses on the effects of spatial quality and angular consistency.

In order to assess the perceptual LFI quality, we need to conduct subjective experiments or build objective quality assessment models. Although subjective evaluation is the most reliable method for measuring LFI quality, it is resource and time consuming and cannot be integrated into the automatic optimization of LFI processing. Therefore, an effective objective LFI quality assessment model is necessary.

\begin{figure}[t]
  \centerline{\includegraphics[width=8cm]{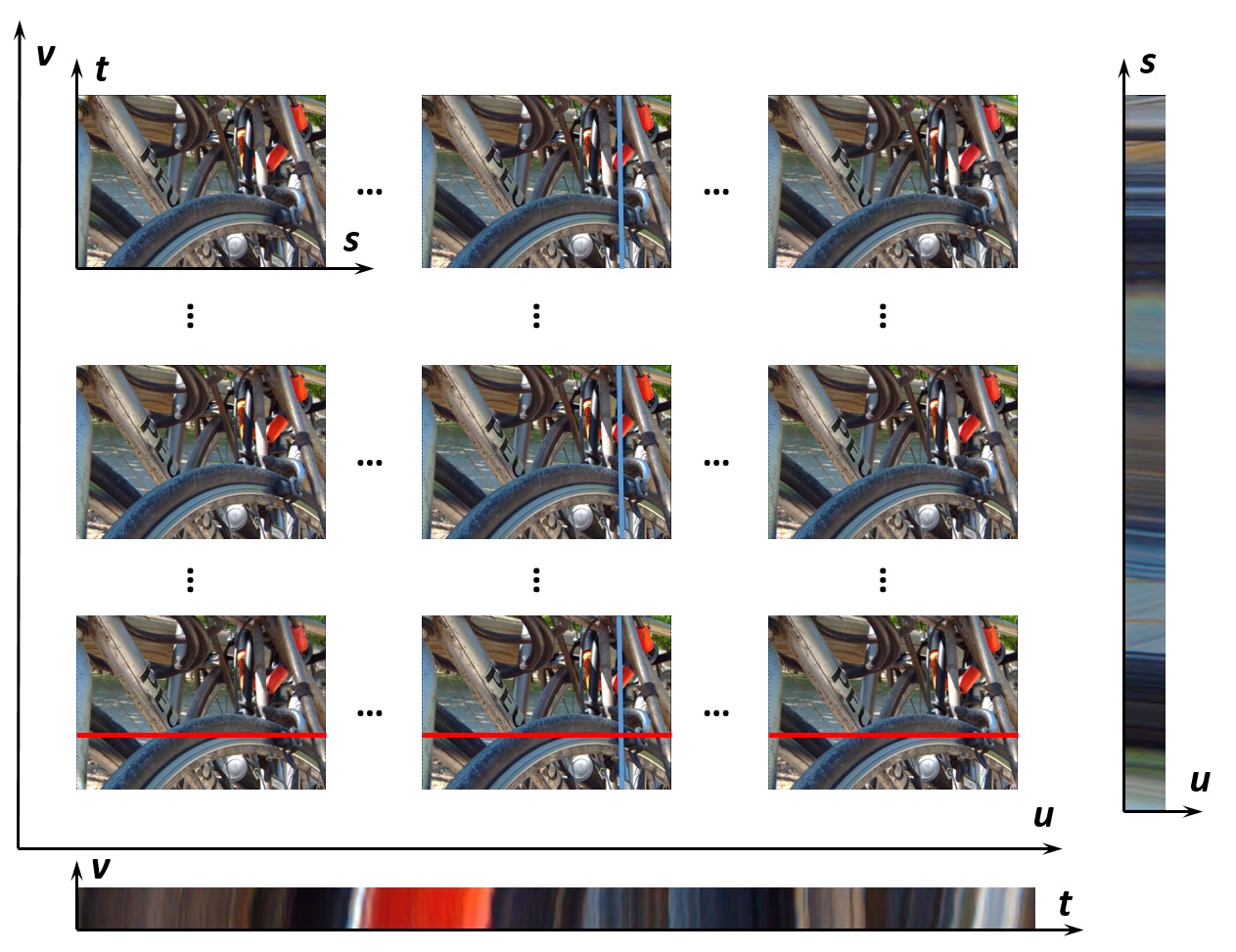}}
  \caption{Light field image captured by Lytro Illum with the corresponding horizontal and vertical epipolar plane image.}
  \centering
\label{fig:SRCs}
\end{figure}

\begin{figure}
\centering
%\begin{tabular}{cc}
\begin{minipage}{0.5\linewidth}
  \centerline{\includegraphics[width=4cm]{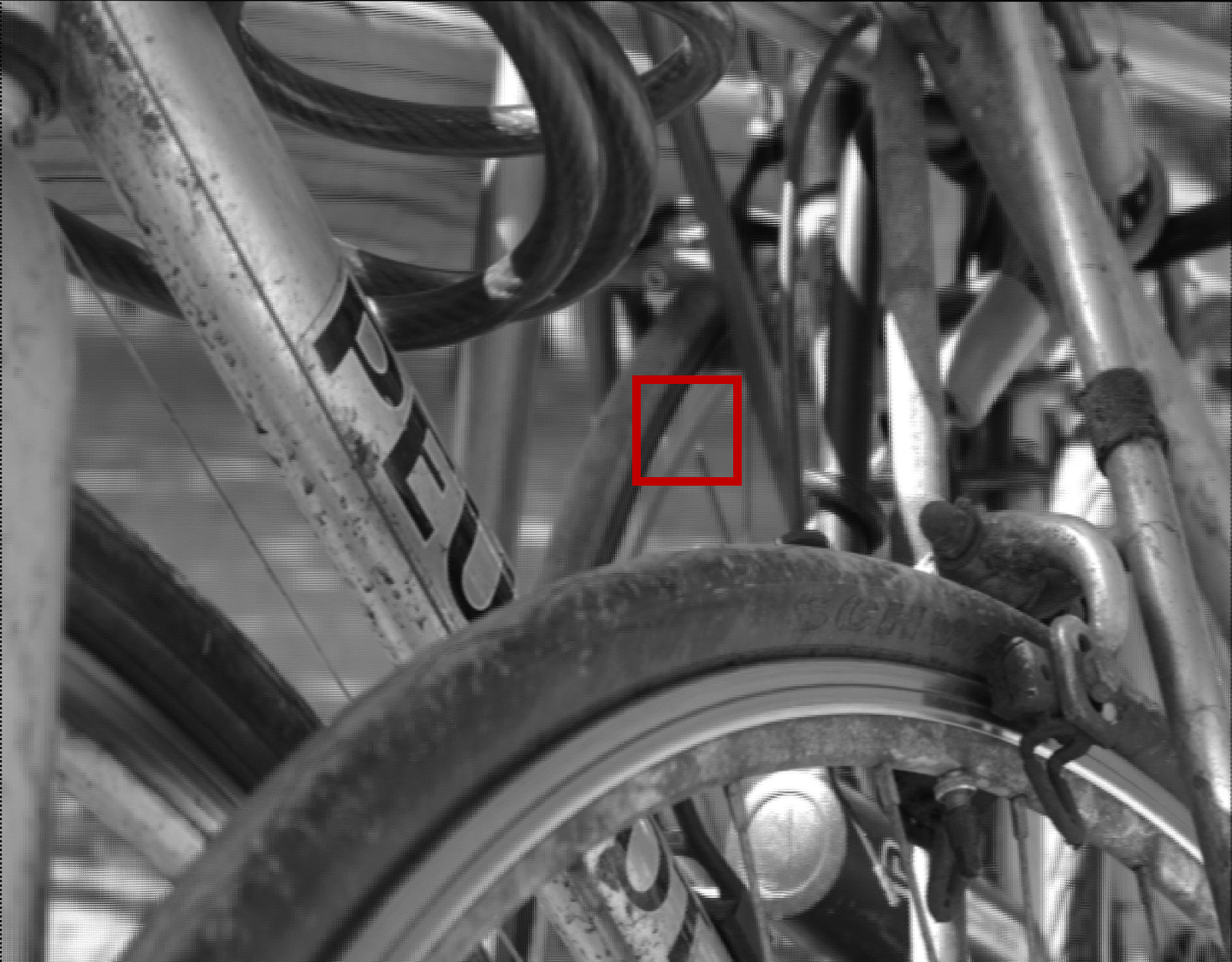}}
  \centerline{(a)}
\end{minipage}
%\hfill
\begin{minipage}{0.4\linewidth}
  \centerline{\includegraphics[width=3.2cm]{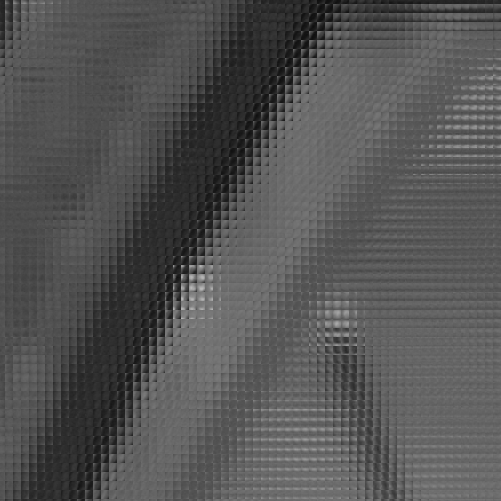}}
  \centerline{(b)}
\end{minipage}
%\hfill
%\end{tabular}
\caption{Lenslet image captured by Lytro Illum. (a) Lenslet image; (b) Amplified regions for the red bounding box of (a).}
\centering
\label{fig:res}
\end{figure}

Generally, according to the availability of original reference information, objective IQA algorithms can be classified as full-reference IQA (FR-IQA), reduced-reference IQA (RR-IQA), and no-reference IQA (NR-IQA). Specifically, the FR-IQA methods utilize complete reference image data and measure the difference between reference and distorted images. The RR-IQA models utilize partial information of the reference image for quality assessment, while the NR-IQA metrics evaluate image quality without original reference images, which is more applicable in most real-world scenarios.
\begin{figure}[!htb]
%\begin{tabular}{cc}
\begin{minipage}{0.55\linewidth}
  \centerline{\includegraphics[width=4.5cm]{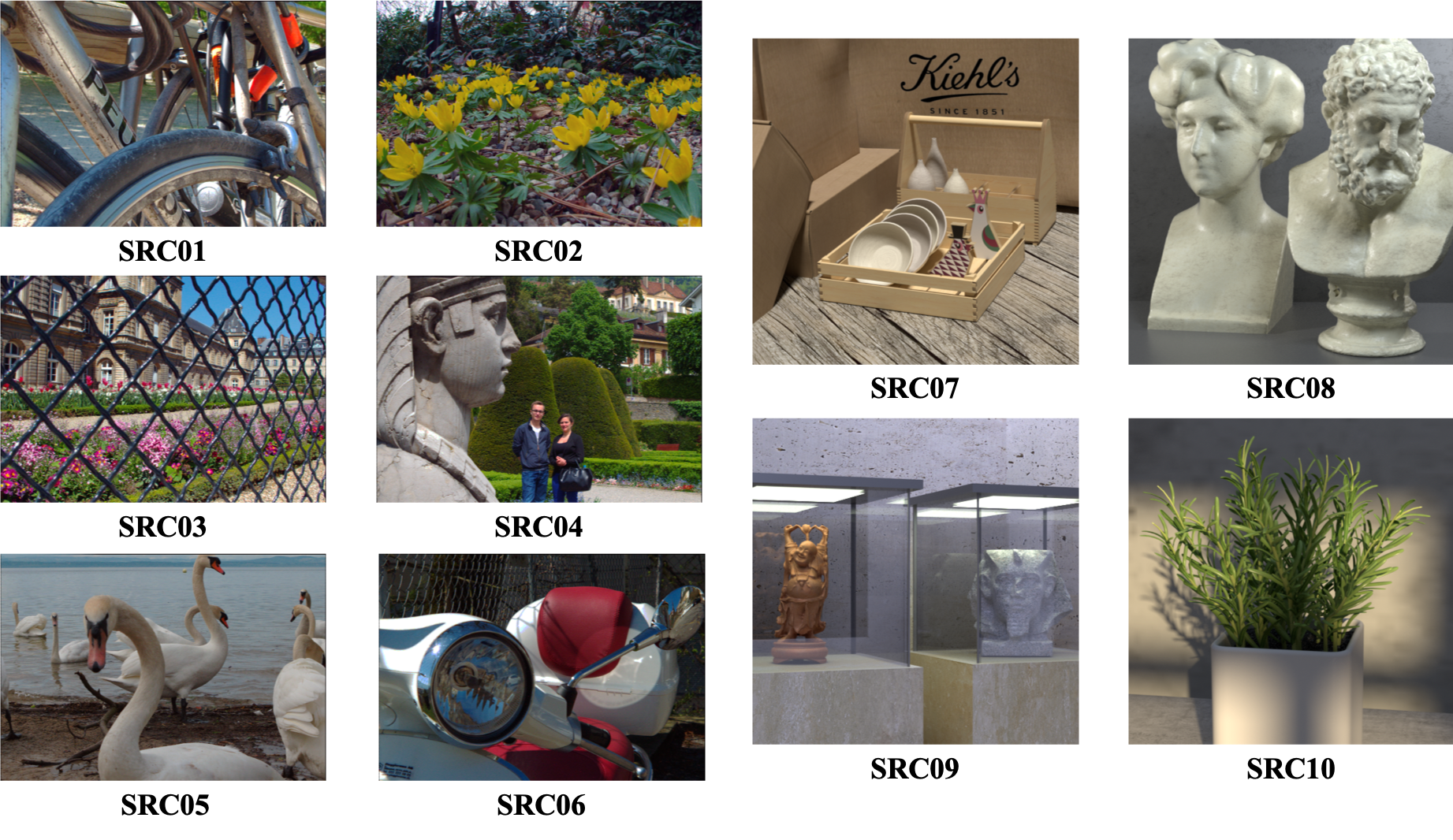}}
  \centerline{(a)}
\end{minipage}
%\hfill
\begin{minipage}{0.44\linewidth}
  \centerline{\includegraphics[width=4.3cm]{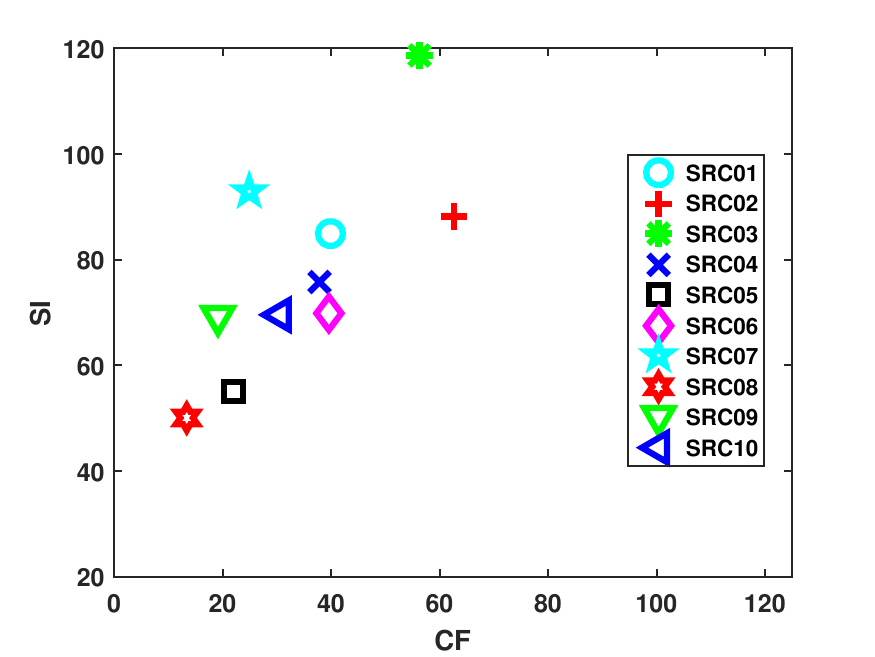}}
  \centerline{(b)}
\end{minipage}
%\vfill
\begin{minipage}{0.55\linewidth}
  \centerline{\includegraphics[width=4.5cm]{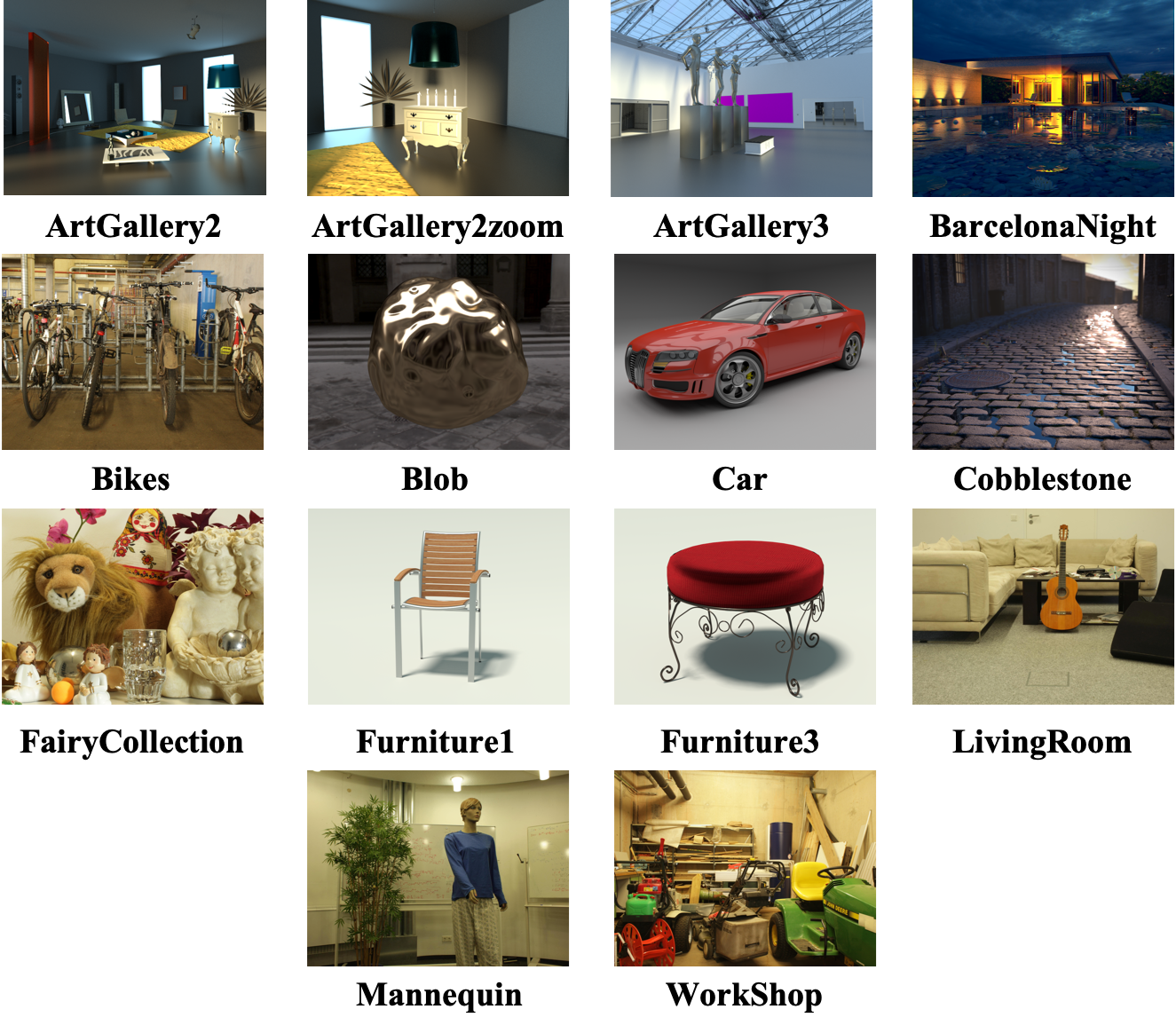}}
  \centerline{(c)}
\end{minipage}
%\hfill
\begin{minipage}{0.44\linewidth}
  \centerline{\includegraphics[width=4.3cm]{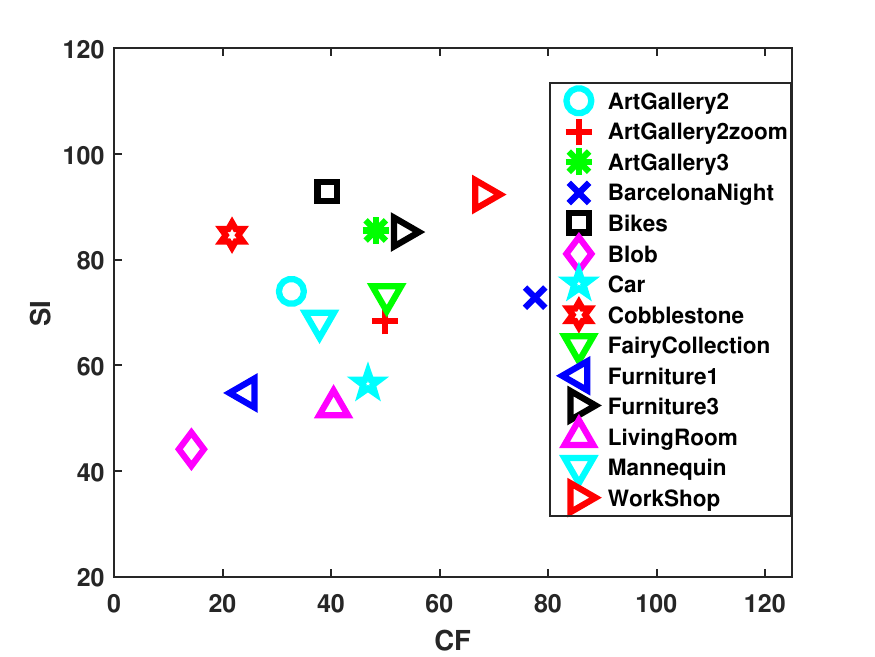}}
  \centerline{(d)}
\end{minipage}
%\hfill
\begin{minipage}{0.55\linewidth}
  \centerline{\includegraphics[width=4.5cm]{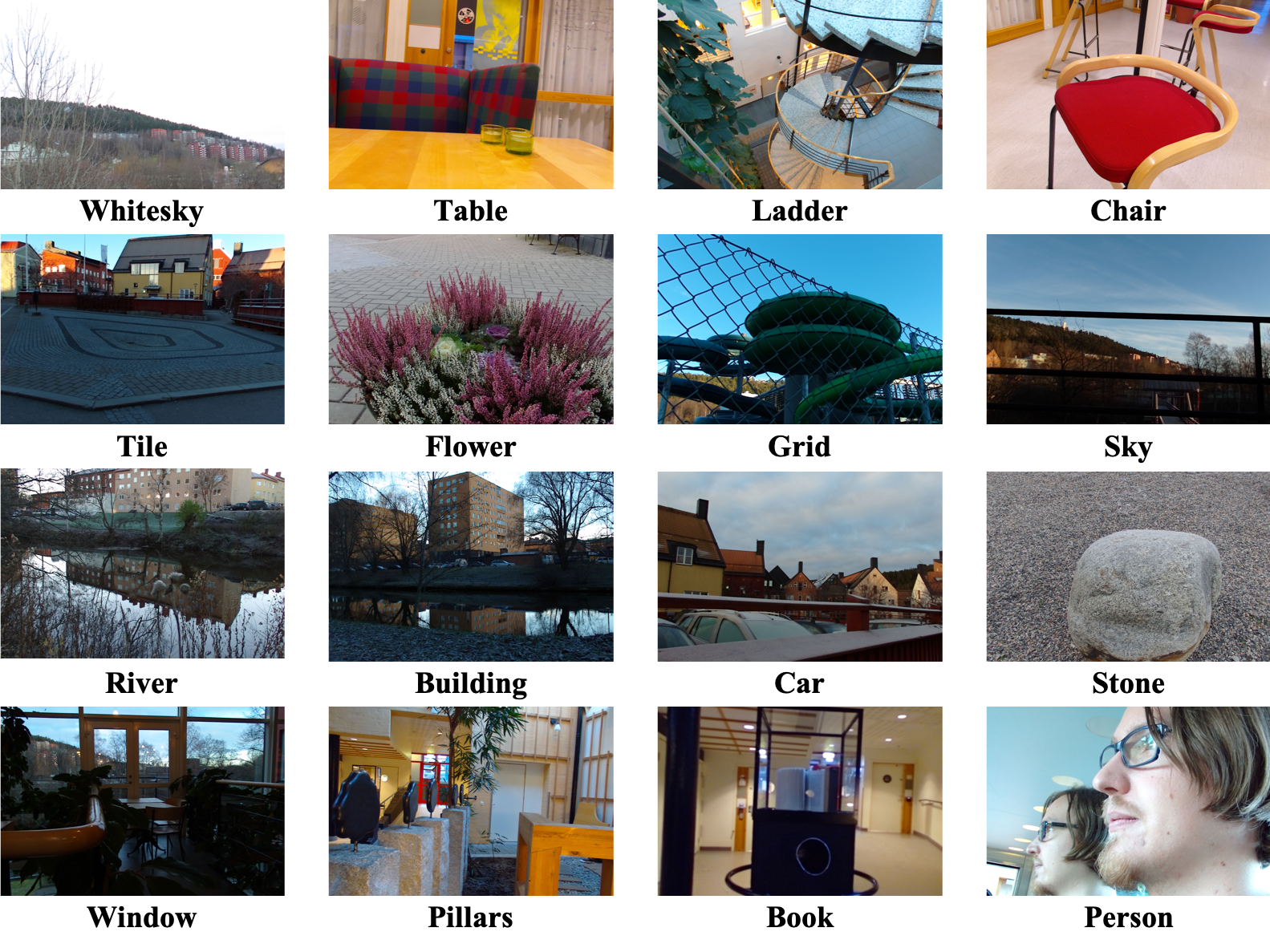}}
  \centerline{(e)}
\end{minipage}
%\hfill
\begin{minipage}{0.44\linewidth}
  \centerline{\includegraphics[width=4.3cm]{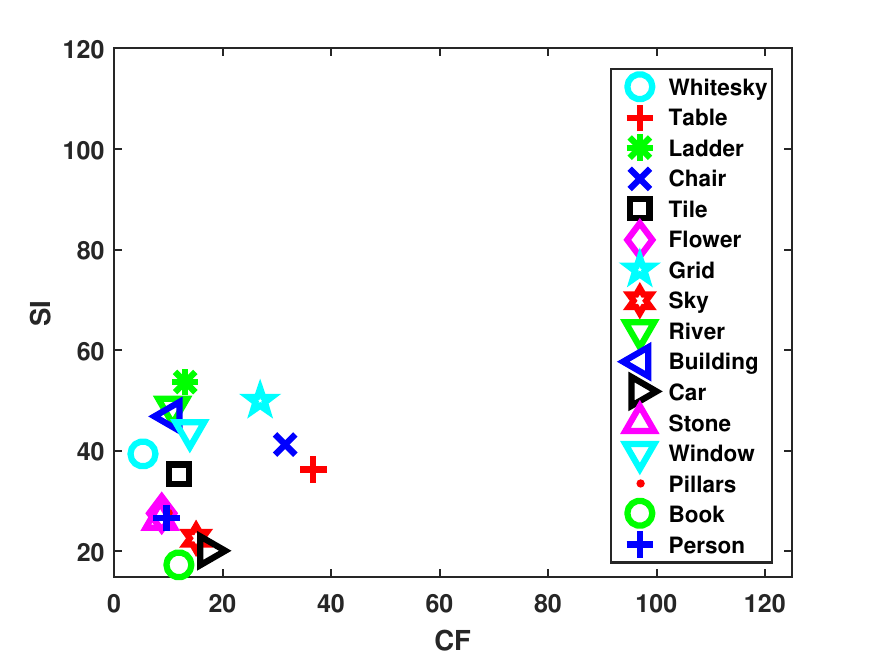}}
  \centerline{(f)}
\end{minipage}
%\hfill
\begin{minipage}{0.55\linewidth}
  \centerline{\includegraphics[width=4.5cm]{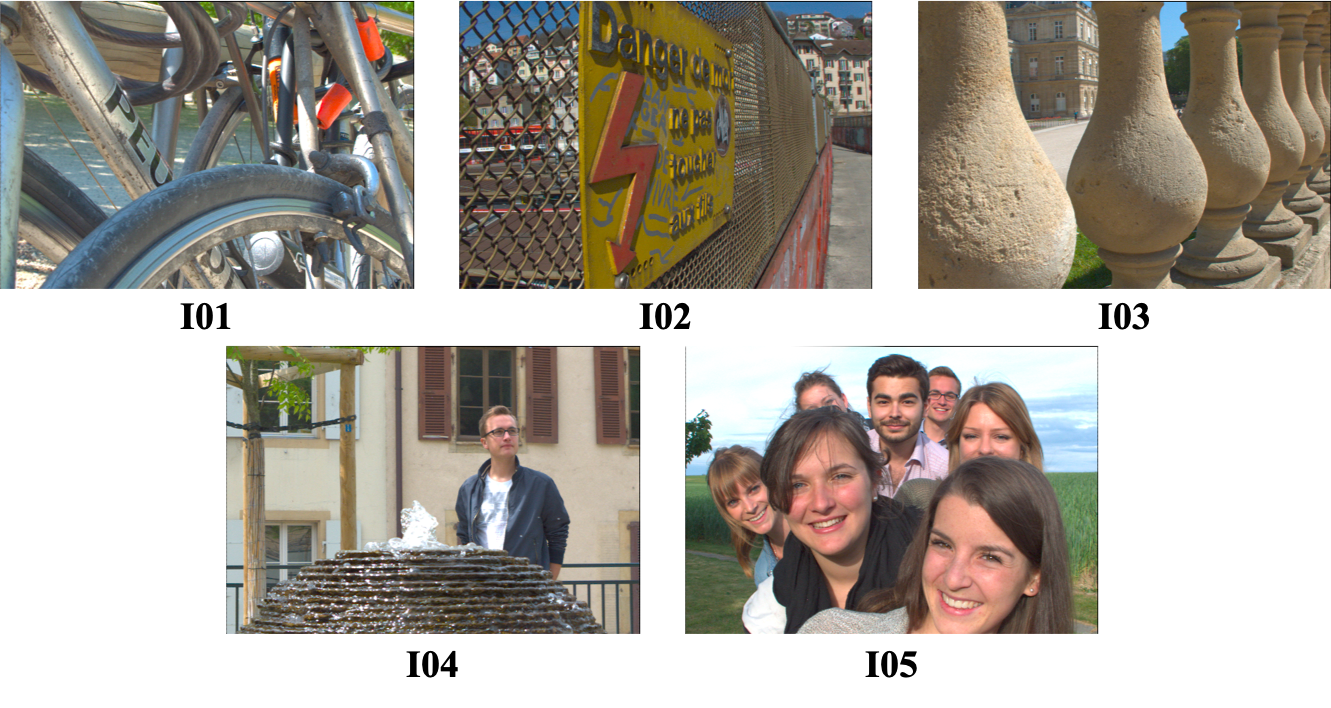}}
  \centerline{(g)}
\end{minipage}
%\hfill
\begin{minipage}{0.44\linewidth}
  \centerline{\includegraphics[width=4.3cm]{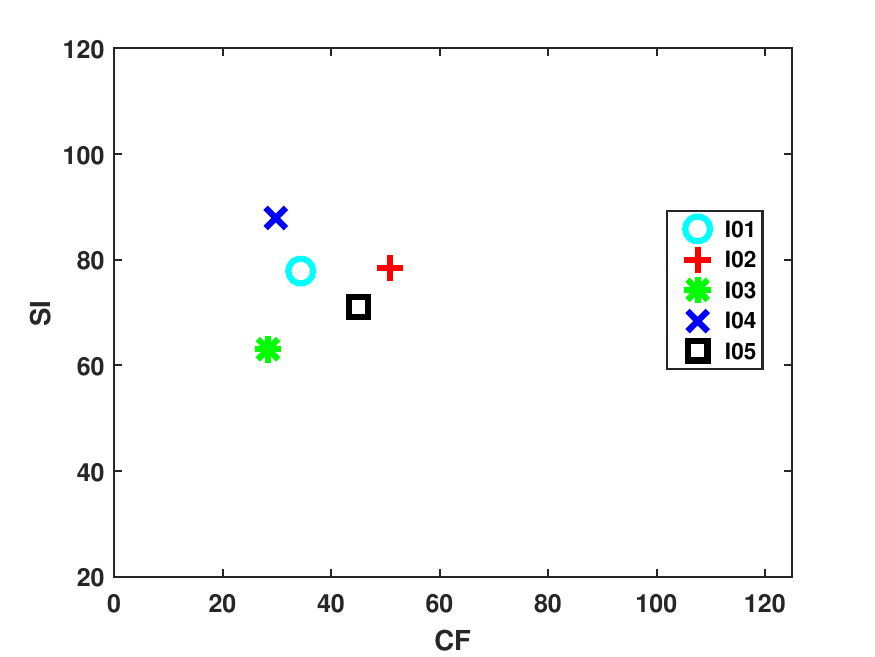}}
  \centerline{(h)}
\end{minipage}
%\hfill
\caption{(a) Illustration of the center view for the selected image contents of source sequences of Win5-LID; (b) Distribution of Spatial perceptual Information (SI) and Colorfulness (CF) of Win5-LID; (c) Illustration of the center view for the selected image contents of source sequences of MPI-LFA; (d) Distribution of SI and CF of MPI-LFA; (e) Illustration of the center view for the selected image contents of source sequences of SMART; (f) Distribution of SI and CF of SMART; (g) Illustration of the center view for the selected image contents of source sequences of VALID; (h) Distribution of SI and CF of VALID.}
\centering
\label{fig:res}
\end{figure}

For 2D IQA, several 2D FR-IQA metrics have been proposed, for example, structural similarity (SSIM) \cite{wang2004image}, multi-scale SSIM (MS-SSIM) \cite{wang2003multiscale}, feature similarity (FSIM) \cite{zhang2011fsim}, information content weighting SSIM (IWSSIM) \cite{wang2011information}, internal generative mechanism (IGM) \cite{wu2013perceptual}, visual saliency-induced index (VSI) \cite{zhang2014vsi} and gradient magnitude similarity deviation (GMSD) \cite{xue2014gradient}. The degree of information loss for the distorted image relative to the reference image is evaluated in information fidelity criterion (IFC) \cite{sheikh2005information} and visual information fidelity (VIF) \cite{sheikh2006image}. The noise quality measure (NQM) \cite{damera2000image} and visual signal-to-noise ratio (VSNR) \cite{chandler2007vsnr} consider the human visual system (HVS) sensitivity to different visual signals and the interaction between different signal components. Moreover, there also exist some 2D RR-IQA metrics, such as \cite{wang2005reduced,wang2011reduced,rehman2012reduced}, and 2D NR-IQA metrics. The distortion identification-based image verity and integrity evaluation (DIIVINE) monitors natural image behaviors based on scene statistics \cite{moorthy2011blind}. The generalized Gaussian density function is modeled by block discrete cosine transform (DCT) coefficients in blind image integrity notator using DCT statistics (BLIINDS-II) \cite{saad2012blind}. The blind/referenceless image spatial quality evaluator (BRISQUE) utilizes scene statistics from spatial domain \cite{mittal2012no}. The natural image quality evaluator (NIQE) extracts image local features and fits a multivariate Gaussian model on the extracted local features \cite{mittal2012making}. A ``bag" of features designed in different color spaces based on human perception are extracted in feature maps-based referenceless image quality evaluation engine (FRIQUEE) \cite{ghadiyaram2017perceptual}. In addition, due to the high dynamic properties of light field content, we also compare with the visual difference predictor for high dynamic range images (HDR-VDP2) \cite{mantiuk2011hdr}.

As for 3D IQA, Yang \textit{et al.} proposed a 3D FR-IQA method based on the average peak signal-to-noise ratio (PSNR) and the absolute difference between left and right views \cite{yang2009objective}. Chen \textit{et al.} proposed a 3D FR-IQA algorithm that models the influence of binocular rivalry \cite{chen2013full}. They also proposed a 3D NR-IQA algorithm that extracts features from cyclopean images, disparity maps, and uncertainty maps \cite{chen2013no}. The S3D integrated quality (SINQ) considers the impact of binocular fusion, rivalry, suppression, and a reverse saliency effect on the distortion perception \cite{liu2017binocular}. The depth information and binocular vision theory are utilized in (BSVQE) \cite{chen2018blind}.

Regarding to multi-view IQA, the morphological pyramid PSNR (MP-PSNR) Full \cite{sandic2015dibr} and MP-PSNR Reduc \cite{sandic2016multi} were proposed. Furthermore, morphological wavelet PSNR (MW-PSNR) adopts morphological wavelet decomposition to evaluate multi-view image quality \cite{sandic2015dibr1}. The 3D synthesized view image quality metric (3DSwIM) is based on the comparison of statistical features extracted from wavelet subbands for original and distorted multi-view images \cite{battisti2015objective}. Gu \textit{et al.} proposed a NR multi-view IQA metric that employs the autoregression based local image description, which is named as AR-plus thresholding (APT) \cite{gu2018model}.

However, the above-mentioned algorithms do not take into account the essential characteristics of the LFI. Specifically, the 2D and 3D algorithms ignore the influence of angular consistency, and the multi-view methods cannot effectively measure the deterioration of spatial quality. Therefore, it is important to design an effective light-field-specific metric.

Although there exist the evaluation methods for light field rendering and visualization \cite{shidanshidi2015estimation,kara2019key}, few objective LFI quality evaluation methods have been proposed.
Specifically, Fang \textit{et al.} proposed a FR LFI-QA method that measures the gradient magnitude similarity of reference and distorted EPIs \cite{fang2018light}.
The RR LF image quality assessment metric (LF-IQM) assumes that depth map quality is closely related to LFIs overall quality, which measures the structural similarity between original and distorted depth maps to predict the LFI quality \cite{8632960}.
However, neither of these methods utilize texture information of SAI, which causes insufficient measurement of LFI spatial quality. Furthermore, LF-IQM performance is significantly affected by depth estimation algorithms.
Additionally, these methods require the use of original LFI information, which limits their application scenarios.
Therefore, a general-purpose NR LFI-QA method that fully considers the factors affecting LFI quality is necessary for real applications.

\begin{figure*}[!htb]
\centering
\includegraphics[width=16cm]{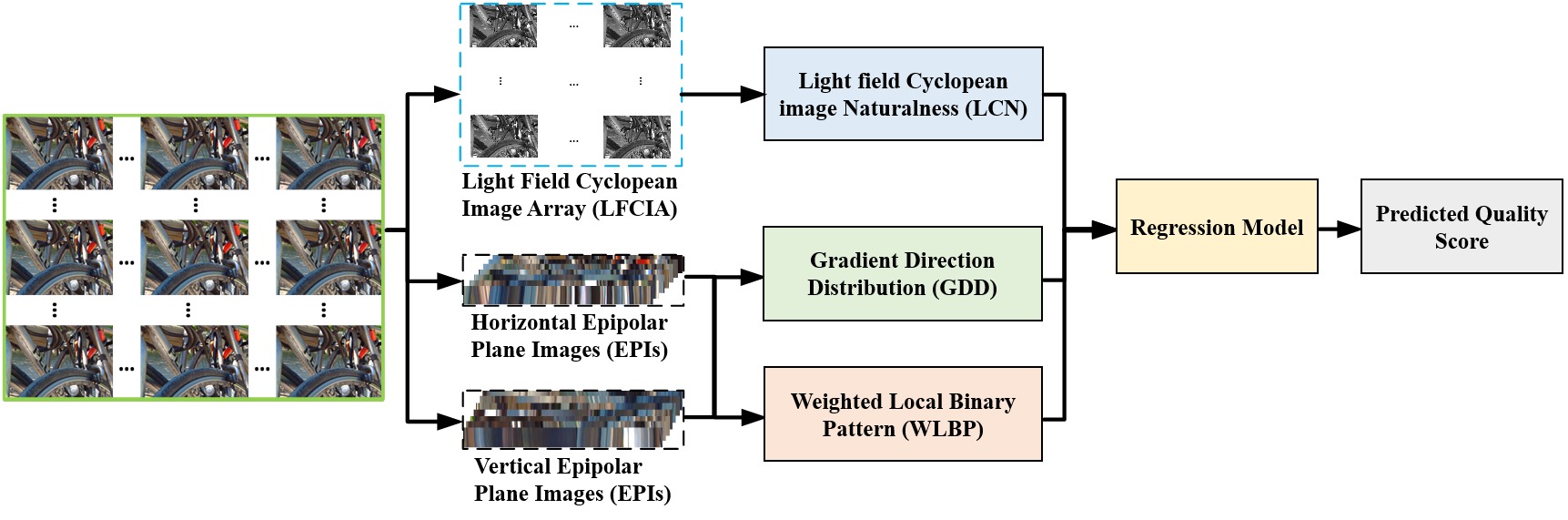}
\caption{Flow diagram of the proposed No-Reference Light Field image Quality Assessment (NR-LFQA) model.}
\centering
\label{fig:SRCs}
\end{figure*}

In this paper, to the best of our knowledge, we propose the first No-Reference Light Field image Quality Assessment (NR-LFQA) scheme.
As the LFI can provide the binocular cue and consists of many SAIs \cite{shi2018light}, we propose Light field Cyclopean image array Naturalness (LCN) to measure the deterioration of spatial quality in LFI, which can take advantage of information from all SAIs and effectively capture the global naturalness distribution of LFI.
Specifically, we first mimic binocular fusion and rivalry to generate Light Field Cyclopean Image Array (LFCIA) and then analyze its naturalness distribution.
In addition to spatial quality, angular consistency is equally important for the LFI perception \cite{shi2018light,adhikarla2017towards}.
Since the EPI contain angle information of the LFI, we extract features for measuring the degradation of angular consistency on EPI including two key aspects:
i) we propose the new Gradient Direction Distribution (GDD) to measure the distribution of EPI gradient direction maps, which can represent the global angular distribution;
ii) we then propose the Weighted Local Binary Pattern (WLBP) descriptor to capture the relationship between different SAIs, which focuses on the local angular consistency degradation.
Finally, these two features can collectively reflect changes in angular consistency.
Our experimental results show that the performance of our proposed NR-LFQA model correlates well with human visual perception and achieves the state-of-the-art performance compared to other objective methods. The software release of NR-LFQA is available online: \url{http://staff.ustc.edu.cn/~chenzhibo/resources.html} for public research usage.

The remainder of this paper is organized as follows. Section II introduces adopted datasets in details. In Section IIII, the proposed model is presented. We then illustrate the experimental results in Section IV. Finally, section V concludes our paper.

\section{Datasets}
We conducted experiments on four publicly available datasets, namely Win5-LID \cite{shi2018light}, MPI-LFA \cite{adhikarla2017towards}, SMART \cite{paudyal2017towards} and VALID \cite{viola2018valid} to examine our proposed NR-LFQA model.

The Win5-LID dataset selects 6 real scenes captured by Lytro illum and 4 synthetic scenes as original images, which cover various Spatial perceptual Information (SI) and Colorfulness (CF) \cite{itu1999subjective}, as shown in Fig. 3. The 220 distorted LFIs were produced by 6 distortion types, including HEVC, JPEG2000 (JPEG), linear interpolation (LN), nearest neighbor interpolation (NN) and two CNN models. In addition to CNN models, each distortion includes 5 intensity levels. The observers were asked to rate 220 LFIs under double-stimulus continuous quality scale (DSCQS) on a 5-point discrete scale. The associated overall mean opinion score (MOS) value is provided for each LFI.

As shown in Fig. 3, the MPI-LFA dataset is composed of 14 pristine LFIs captured by the TSC system, covering various SI and CF. Moreover, 336 distorted LFIs were produced based on 6 distortion types (i.e. HEVC, DQ, OPT, LINEAR, NN and GAUSS), with 7 degradation levels for each distortion type. To quantify LFI quality, pair-wise comparison (PC) method with a two-alternative-forced-choice is performed and just-objectionable-differences (JOD) is provided, which is more similar to difference-mean-opinion-score (DMOS). The lower value indicates the worse quality.

The SMART dataset contains 16 original light field images and 256 distorted sequences are obtained by introducing 4 compression distortions (i.e. HEVC intra, JPEG, JPEG2000 and SSDC). The PC method is selected to collect the subjective ratings for each image and the Bradley-Terry (BT) scores are provided. Higher BT score indicates higher preference rate.

The VALID dataset consists of 5 reference LFIs with 140 distorted LFIs spanning 5 compression solutions. The VALID includes both 8bit and 10bit LFIs. The comparison-based adjectival categorical judgement methodology is adopted for 8bit images, while double stimulus impairment scale (DSIS) is performed for 10bit images. Moreover, the MOS values are provided for the LFIs.

\section{Proposed Method}
The block diagram of the proposed model is shown in Fig. 4. First, we analyze the naturalness distribution of LFCIA, which is generated by simulating binocular fusion and binocular rivalry. To measure the degradation of angular consistency, GDD is then computed to analyze the direction distribution of EPI gradient direction maps. The WLBP descriptor is proposed to measure the relationship between different SAIs. Finally, regression model is used to predict LFI quality.

\begin{figure*}[!htb]
\centering
\begin{minipage}{0.31\linewidth}
  \centerline{\includegraphics[width=6cm]{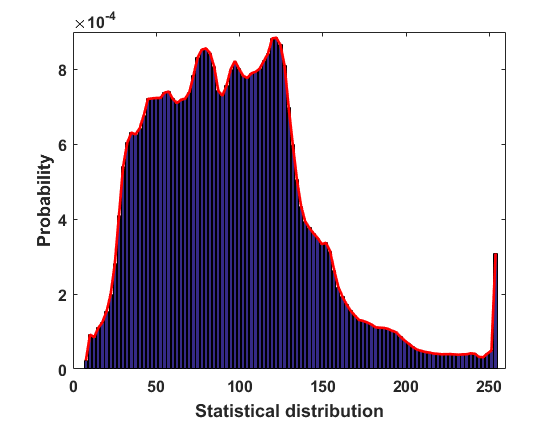}}
  \centerline{(a)}
\end{minipage}
%\hfill
\begin{minipage}{0.31\linewidth}
  \centerline{\includegraphics[width=6cm]{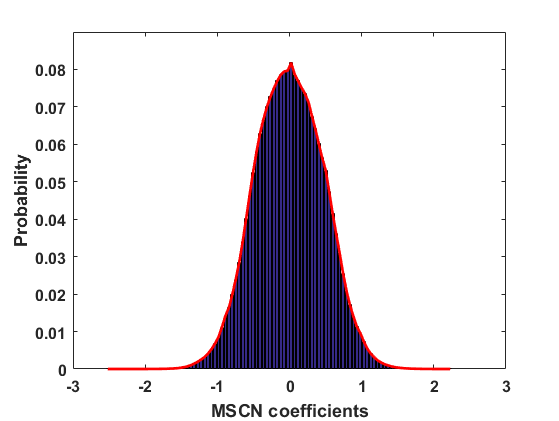}}
  \centerline{(b)}
\end{minipage}
%\hfill
\begin{minipage}{0.31\linewidth}
  \centerline{\includegraphics[width=6cm]{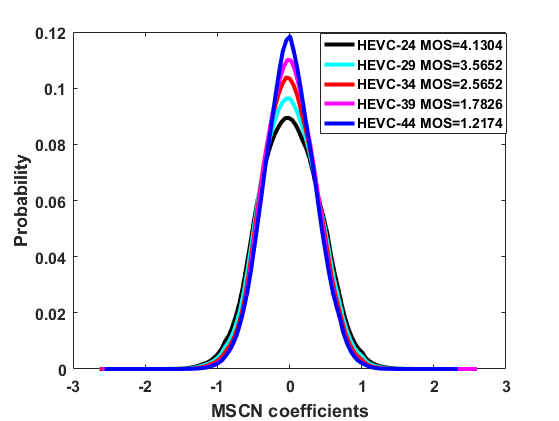}}
  \centerline{(c)}
\end{minipage}
%\end{tabular}
\caption{Statistical distribution of Light Field Cyclopean Image Array (LFCIA) and Mean Subtracted and Contrast Normalized (MSCN) coefficients. (a) Statistical distribution of LFCIA; (b) MSCN coefficients of LFCIA; (c) MSCN coefficients for different HEVC compression levels.}
\centering
\label{fig:res}
\end{figure*}

\subsection{Light field Cyclopean image array Naturalness (LCN)}
In general, as the LFI perceptual quality is decided by the HVS, it is reasonable to quantify the LFI quality by characterizing the human perception process.
Considering that the LFI can provide binocular perception, we utilize the binocular fusion and binocular rivalry theory to evaluate the spatial quality of LFI.
In specific, the disparity between the left and right views of the LFI is small, so most scenes are guaranteed in the comfort zone and binocular fusion occurs \cite{shibata2011zone}.
When the perceived contents to the left and right eyes are significantly different, the failures of binocular fusion will lead to binocular rivalry \cite{steinman2000foundations}.

To simulate the human perception process, the cyclopean perception theory provides us with an executable strategy to achieve this goal \cite{julesz1971foundations}.
When a stereoscopic image pair is presented, a cyclopean image is formed in the mind of observers.
The generated cyclopean image contains both left and right view information and takes into account the impact of binocular fusion and binocular rivalry characteristics, so it can effectively reflect the perceived image quality \cite{chen2013full,liu2017binocular}.

Specifically, a cyclopean image is synthesized from the stereoscopic image, disparity map, and spatial activity map. Here, the horizontal adjacent SAIs of the LFI are regraded as a left view and a right view separately. We define the left view as $I_{u,v}(s,t)$, where $(s,t)$ represents the SAI spatial coordinates and $(u,v) \in \{U,V\}$ indicates the angular coordinate of the LFI. The LFCIA with angular resolution of $U\times (V-1)$ can be synthesized by the following model:
\begin{equation}
\begin{aligned}
C_{u,v}(s,t)= W_{u,v}(s,t) \times I_{u,v}(s,t) + \\
 W_{u+1,v}(s,t) \times I_{u+1,v}((s,t)+d_{s,t}),
\end{aligned}
\end{equation}
where $C_{u,v}$ represents a sub-cyclopean image at angular coordinate $(u,v)$ and
$d_{s,t}$ is the horizontal disparity between $I_{u,v}$ and $I_{u+1,v}$ at spatial coordinate $(s,t)$. The disparity map $d$ is generated by using a simple stereo disparity estimation algorithm, which utilizes the SSIM index as a matching criterion \cite{chen2013full}. The weights $W_{u,v}$ and $W_{u+1,v}$ are computed by the following formulas:

\begin{small}
\begin{equation}
\begin{aligned}
W_{u,v}(s,t) = \frac{\varepsilon[S_{u,v}(s,t)]+A_1}{\varepsilon[S_{u,v}(s,t)]+\varepsilon[S_{u+1,v}((s,t)+d_{s,t})]+A_1},
%W_{u,v}(i)=\frac{\{\varepsilon[S_{u,v}(i)]+A_2\}{\{\varepsilon[S_{u,v}(i)]+A_2\}+\{\varepsilon[S_{u,v+1}(i+d_i)]+A_2\}}
\end{aligned}
\end{equation}
\end{small}

\begin{small}
\begin{equation}
\begin{aligned}
W_{u+1,v}(s,t) = \frac{\varepsilon[S_{u+1,v}((s,t)+d_{s,t})]+A_1}{\varepsilon[S_{u,v}(s,t)]+\varepsilon[S_{u,v+1}((s,t)+d_{s,t})]+A_1},
%W_{u,v}(i)=\frac{\{\varepsilon[S_{u,v}(i)]+A_2\}{\{\varepsilon[S_{u,v}(i)]+A_2\}+\{\varepsilon[S_{u,v+1}(i+d_i)]+A_2\}}
\end{aligned}
\end{equation}
\end{small}

\noindent{where $A_1$ is the small value to guarantee stability. $S_{u,v}(s,t)$ is $N \times N$ region pixels centered on $(s,t)$ and $\varepsilon[S_{u,v}(s,t)]$ is the spatial activity within $S_{u,v}(s,t)$. We then obtain the spatial activity map as follows:}
\begin{equation}
\varepsilon[S_{u,v}(s,t)]=\log_{2}[var_{u,v}^2(s,t)+A_2],
\end{equation}
where $var_{u,v}(s,t)$ is the variance of $S_{u,v}(s,t)$ and unit item $A_2$ guarantee the activity is positive. Similarly, we define quantities $S_{u+1,v}(s,t)$, $var_{u+1,v}(s,t)$ and $\varepsilon[S_{u+1,v}(s,t)]$ on the $I_{u+1,v}(s,t)$.

After obtaining the LFCIA, the locally Mean Subtracted and Contrast Normalized (MSCN) coefficients are utilized to measure their naturalness, which have been successfully employed for image processing tasks and can be used to model the contrast-gain masking process in early human vision \cite{mittal2012no,carandini1997linearity}. For each sub-cyclopean image, its MSCN coefficients can be calculated by:
\begin{equation}
\widehat{C}_{u,v}(s,t)=\frac{C_{u,v}(s,t)-\mu_{u,v}(s,t)}{\sigma_{u,v}(s,t)+1},
\end{equation}
where $\widehat{C}_{u,v}(s,t)$ and $C_{u,v}(s,t)$ represent the MSCN coefficients image and sub-cyclopean image values at spatial position $(s,t)$ respectively. $\mu_{u,v}(s,t)$ and $\sigma_{u,v}(s,t)$  stands for the local mean and standard deviation in a local patch centered at $(s,t)$. They are respectively calculated as:
\begin{equation}
\mu_{u,v}(s,t)= \sum\limits_{k=-K}^{K} \sum\limits_{l=-L}^{L} z_{k,l} C_{u,v}^{k,l}(s,t),
\end{equation}

\begin{equation}
\sigma_{u,v}(s,t) = \sqrt{\sum\limits_{k=-K}^{K}\sum\limits_{l=-L}^{L} z_{k,l} (C_{u,v}^{k,l}(s,t)-\mu_{u,v}(s,t))^2},
\end{equation}
where $z=\{z_{k,l}|k=-K,...,K,l=-L,...,L\}$ denotes a 2D circularly-symmetric gaussian weighting function sampled out 3 standard deviations and rescaled to unit volume. We set $K=L=3$ in our implement.

To measure the spatial quality of the LFI, we consider the naturalness distribution of LFCIA. In specific, the MSCN coefficients of all sub-cyclopean images are superimposed. As shown in Fig. 5(a) and Fig. 5(b), the distribution of MSCN coefficients is significantly different from the statistical distribution of sub-cyclopean images and approximates the Gaussian distribution. Fig. 5(c) illustrates the MSCN coefficients distributions of LFIs with different high efficiency video coding (HEVC) compression levels, the results of which represent that the MSCN coefficients distributions are very indicative when the LFI suffer from distortions. Here the sample of LFI is selected from the Win5-LID dataset \cite{shi2018light}.
Then, we utilize the zero-mean asymmetric generalized gaussian distribution (AGGD) model to qualify MSCN coefficients distribution, which can fit the distribution by:

\begin{figure}[!htb]
	\centering
	\includegraphics[width=8.5cm]{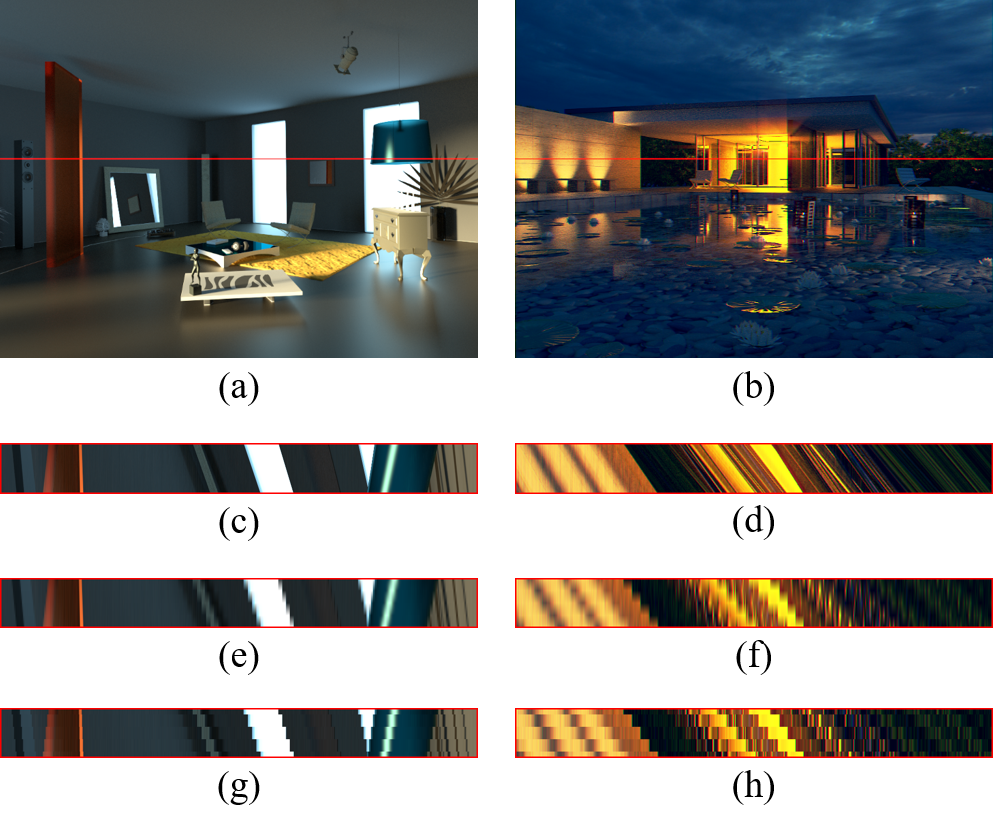}
	\caption{Horizontal EPI for a undistorted LFI and its various distorted versions. The samples are selected from the MPI-LFI dataset \cite{adhikarla2017towards}. (a-b) Center view of the LFIs and the red line indicates the row position at which the EPI is acquired; (c-d) Original EPIs; (e-f) Distorted EPIs with LINEAR distortion; (g-h) Distorted EPIs with NN distortion.}
    \centering
	\label{fig_epicmp}
\end{figure}

\begin{figure}[!htb]
	\centering
	\includegraphics[width=8.5cm]{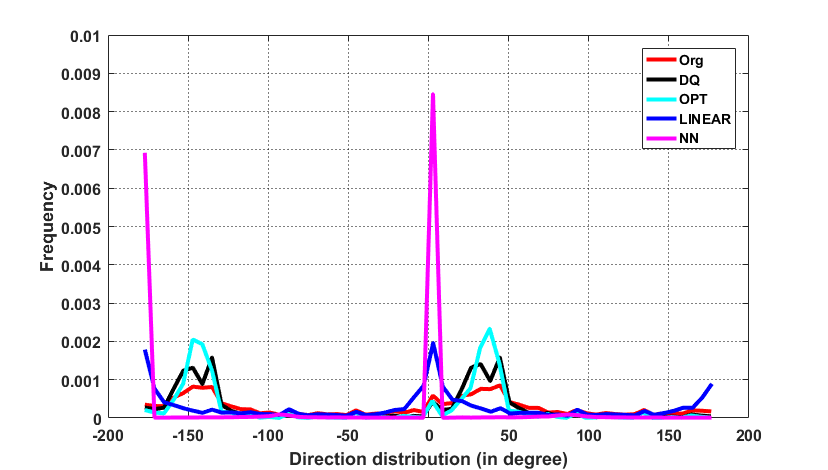}
	\caption{Gradient direction distribution for a undistorted LFI and its various distorted types.}
    \centering
	\label{fig:SRCs}
\end{figure}

\begin{small}
\begin{equation}
f(x;\alpha,\sigma_l^2,\sigma_r^2)=\left\{
\begin{aligned}
\frac{\alpha}{(\beta_l+\beta_r)\Gamma(\frac{1}{\alpha})}exp(-(\frac{-x}{\beta_l})^\alpha) & & x<0 \\
\frac{\alpha}{(\beta_l+\beta_r)\Gamma(\frac{1}{\alpha})}exp(-(\frac{-x}{\beta_r})^\alpha) & & x \geqslant 0,
\end{aligned}
\right.
\end{equation}
\end{small}

\noindent{where}
\begin{equation}
\beta_l=\sigma_l\sqrt{\frac{\Gamma(\frac{1}{\alpha})}{\Gamma(\frac{3}{\alpha})}}  \quad and \quad \beta_r=\sigma_r\sqrt{\frac{\Gamma(\frac{1}{\alpha})}{\Gamma(\frac{3}{\alpha})}},
\end{equation}
and $\alpha$ is the shape parameter controlling the shape of the statistic distribution, while $\sigma_l$, $\sigma_r$ are the scale parameters of the left and right sides respectively. If $\sigma_l=\sigma_r$, AGGD can become generalized gaussian distribution (GGD) model. In addition, we utilize aforementioned three parameters to compute $\eta$ as another feature:

\begin{equation}
\eta=(\beta_r-\beta_l)\frac{\Gamma(\frac{2}{\alpha})}{\Gamma(\frac{1}{\alpha})}.
\end{equation}

We further compute kurtosis and skewness as supplementary features. In addition, the SAIs are downsampled by a factor of 2, which has been proven to improve the correlation between model prediction and subjective assessment \cite{mittal2012no}.
Finally, LFCIA naturalness $F_{LCN}$ is obtained.

\subsection{Global Direction Distribution (GDD)}
\begin{figure*}[!htb]
\centering
%\begin{tabular}{cc}
\begin{minipage}{0.23\linewidth}
  \centerline{\includegraphics[width=4cm]{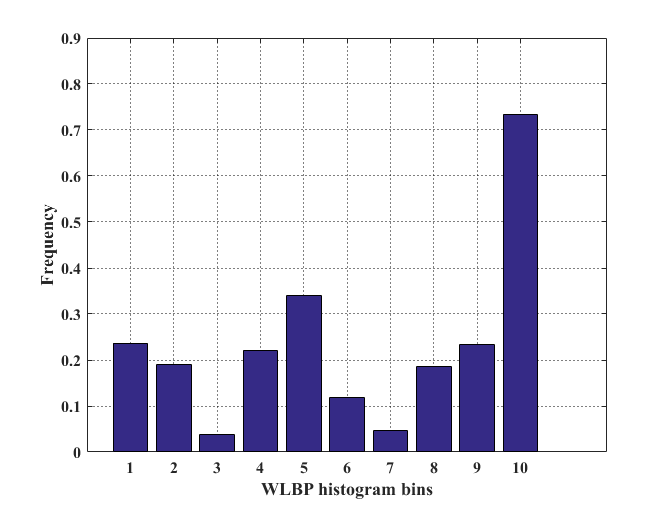}}
  \centerline{(a)}
\end{minipage}
%\vfill
\begin{minipage}{0.23\linewidth}
  \centerline{\includegraphics[width=4cm]{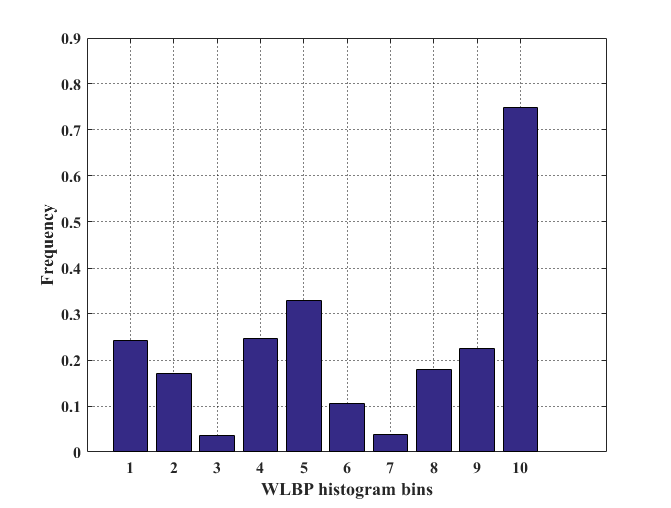}}
  \centerline{(b)}
\end{minipage}
%\vfill
\begin{minipage}{0.23\linewidth}
  \centerline{\includegraphics[width=4cm]{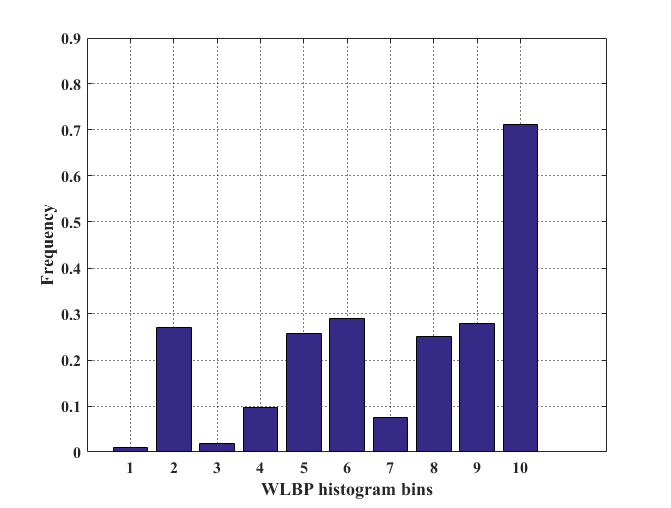}}
  \centerline{(c)}
\end{minipage}
%\vfill
\begin{minipage}{0.23\linewidth}
  \centerline{\includegraphics[width=4cm]{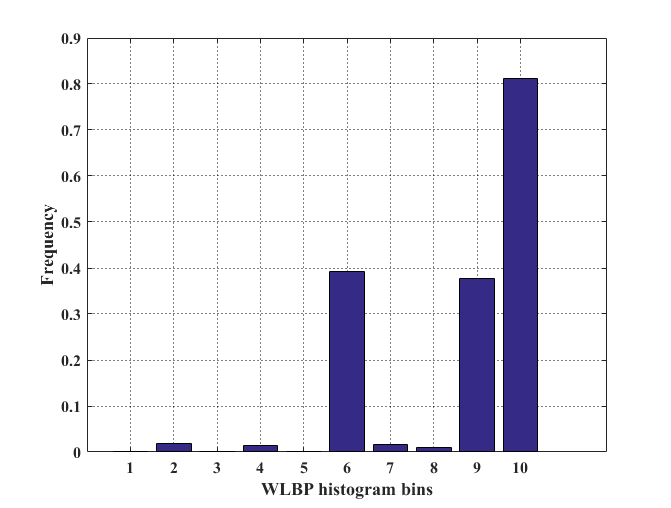}}
  \centerline{(d)}
\end{minipage}
\caption{Illustration of the statistical distribution of WLBP with different distortion types from MPI-LFA \cite{adhikarla2017towards}. Here we set $R=1$ and $P=8$. (a) Quantized depth maps (DQ). (b) Optical flow estimation (OPT). (c) LINEAR interpolation. (d) Nearest Neighbor (NN) interpolation.}
\centering
\label{fig:res}
\end{figure*}

\begin{figure*}[!htb]
\centering
%\begin{tabular}{cc}
\begin{minipage}{0.23\linewidth}
  \centerline{\includegraphics[width=4cm]{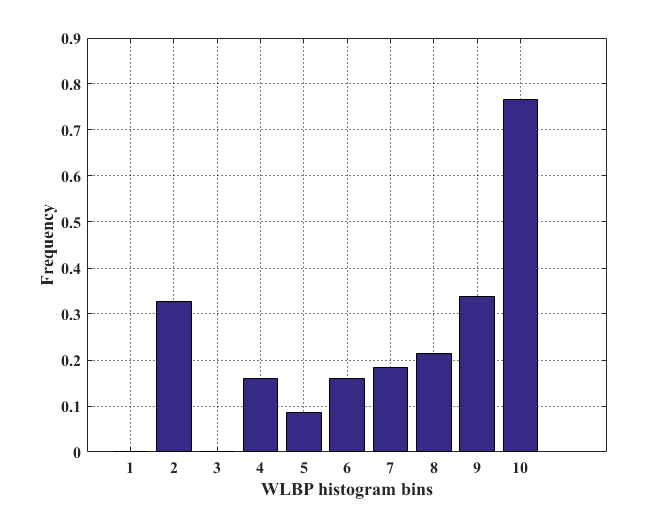}}
  \centerline{(a)}
\end{minipage}
%\vfill
\begin{minipage}{0.23\linewidth}
  \centerline{\includegraphics[width=4cm]{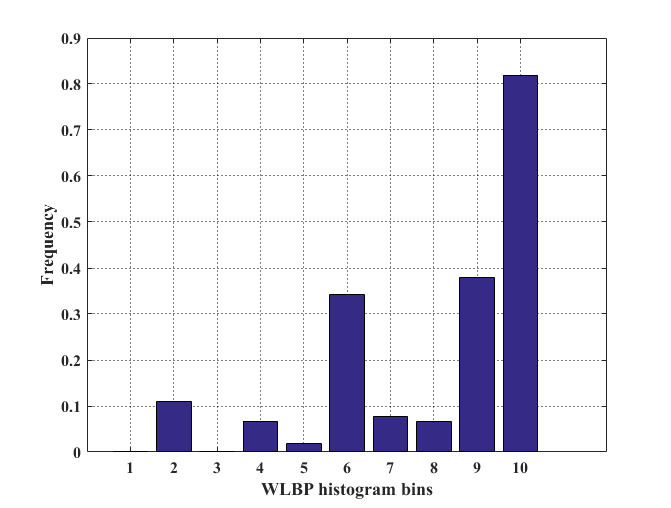}}
  \centerline{(b)}
\end{minipage}
%\vfill
\begin{minipage}{0.23\linewidth}
  \centerline{\includegraphics[width=4cm]{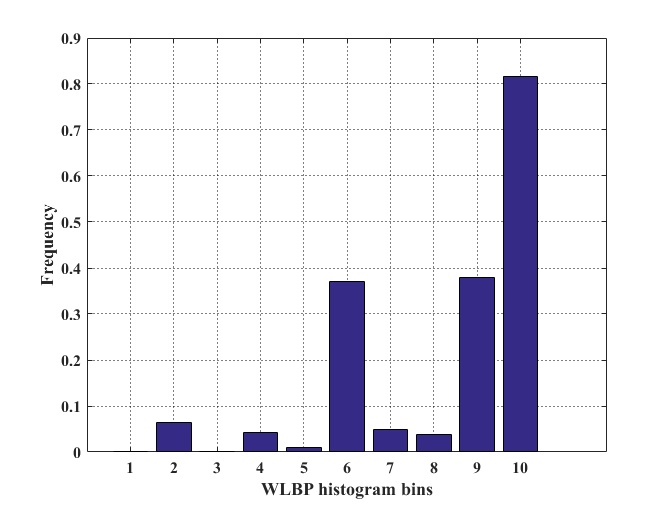}}
  \centerline{(c)}
\end{minipage}
%\vfill
\begin{minipage}{0.23\linewidth}
  \centerline{\includegraphics[width=4cm]{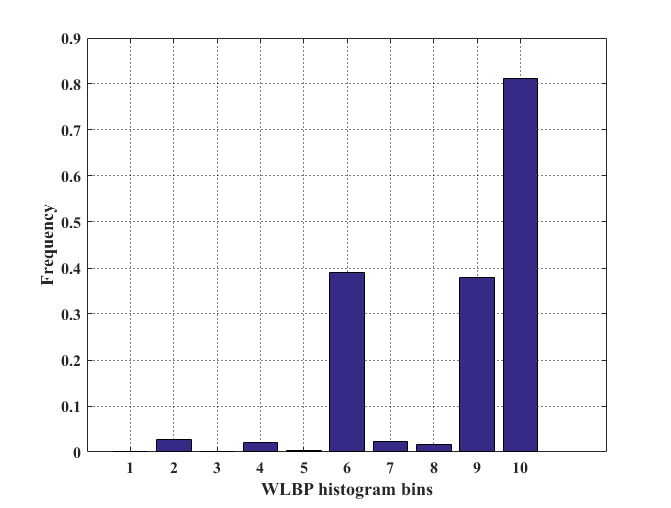}}
  \centerline{(d)}
\end{minipage}
\caption{Illustration of the statistical distribution of WLBP with different NN distortion levels from MPI-LFA \cite{adhikarla2017towards}. Here we set $R=1$ and $P=8$. (a) Level 1; (b) Level 2; (c) Level 3; (d) Level 4.}
\centering
\label{fig:res}
\end{figure*}

The LFI quality is affected by spatial quality and angular consistency.
Usually, angular reconstruction operations, such as interpolation, will break angular consistency.
To measure the deterioration of angular consistency, extracting features from the EPI is an executable strategy because it contains the angular information of the LFI.

Traditionally, the slopes of lines in the EPI reflect the depth of the scene captured by light field. Many LFI processing tasks, such as super resolution and depth map estimation, benefit from this particular property \cite{wu2017light,wu2017light_epi}. As shown in Fig. \ref{fig_epicmp}(c-d), the slopes of lines in the undistorted EPI can effectively reflect the depth and disparity information in Fig. \ref{fig_epicmp}(a-b). Here the LFIs are from MPI-LFA dataset \cite{adhikarla2017towards}.

However, angular distortion damages existing structures and significantly changes the distribution of the slopes of lines in the EPI, as presented in Fig. \ref{fig_epicmp}(e-h). Specifically, we find that the distorted EPIs with the same distortion type have a similar distribution, indicating that the angular distortion is not sensitive to the depth and content of the original LFI. Therefore, the slopes of lines in distorted EPI can be utilized to capture the degradation of angular distortion of LFI.

%As shown in Fig. 5, the EPIs of various reconstruction distortion types is compared with pristine LFI's EPI. The LFIs are from MPI-LFA database \cite{adhikarla2017towards}.
%It can be seen that the original EPI consists of a large number of slopes of lines that reflect the range of disparity \cite{wu2017light}, and we find that the continuity of the slopes of lines can reflect the angular information.
%Therefore, we assume that the direction of lines on the EPI can be used to measure the degradation of angular consistency.
%As presented in Fig. 5, the lines direction of the distorted EPIs have changed significantly compared to the original EPI, especially LINEAR interpolation and NN interpolation. This phenomenon proves that our assumption is feasible.

To extract the Gradient Direction Distribution (GDD) features, we first compute the gradient direction maps of EPIs and then analyze the distribution of the gradient direction maps to obtain the features.
We define the vertical EPI and horizontal EPI as $E_{u^*,s^*}(v,t)$ and $E_{v^*,t^*}(u,s)$ respectively, where $u^*,s^*$ and $v^*,t^*$ represent the fixed coordinates. Then we analyze the direction distribution of EPI by calculating the gradient direction maps of EPI:

\begin{equation}
G_{u^*,s^*}=atan2(-Ey_{u^*,s^*},Ex_{u^*,s^*})*\frac{180}{\pi},
\end{equation}
\noindent{where}
\begin{equation}
Ex_{u^*,s^*}= E_{u^*,s^*} \otimes h_x \quad and \quad Ey_{u^*,s^*}=E_{u^*,s^*} \otimes h_y,
\end{equation}

\begin{equation}
h_x={
\left[ \begin{array}{ccc}
-1 & 0 & 1 \\
-2 & 0 & 2 \\
-1 & 0 & 1
\end{array}
\right ]}
, \quad
h_y={
\left[ \begin{array}{ccc}
-1 & -2 & -1 \\
0 & 0 & 0 \\
1 & 2 & 1
\end{array}
\right ]}.
\end{equation}

Similar to the computation process of $G_{u^*,s^*}$, we can obtain the gradient direction maps $G_{v^*,t^*}$ of horizontal EPI. We then quantify the gradient direction maps to 360 bins, i.e. from $-180^\circ$ to $179^\circ$. As shown in Fig. 7, various distortion types exert systematically different influences on the EPIs. It can be seen that the original EPI has a direction histogram concentrated on about $-150^\circ$ and $30^\circ$, while the distorted EPIs manifest different distributions.
Specifically, the nearest neighbor (NN) and LINEAR interpolation distortions make the EPIs appear stepped, and their direction is mainly concentrated at $-180^\circ$ and $0^\circ$. The optical flow estimation (OPT) and quantized depth maps (DQ) distortions show higher peaks at $-150^\circ$ and $30^\circ$. Overall, gradient direction distribution is effective for measuring the degradation of angular consistency.

Finally, we calculate the mean value, entropy, skewness and kurtosis of the $G_{u^*,s^*}$ and $G_{v^*,t^*}$, respectively. Since the LFI contains many EPIs, we average the mean value, entropy, skewness and kurtosis of all EPIs to obtain the feature $F_{GDD}$.

\subsection{Weighted Local Binary Pattern (WLBP)}
\renewcommand\arraystretch{1.3}

\begin{table*}[!htb]

	\centering

	\scriptsize

	\caption{Performance Comparison on Win5-LID, MPI-LFA, and SMART Datasets.}

	\begin{tabular}{c|c|cccc|cccc|cccc}

		\hline

		\multicolumn{1}{c|}{} & \multicolumn{1}{c|}{} & \multicolumn{4}{c|}{\textbf{Win5-LID}}       & \multicolumn{4}{c|}{\textbf{MPI-LFA}} & \multicolumn{4}{c}{\textbf{SMART}} \\ \hline

		\textbf{Type}                            & \textbf{Metrics}       & \textbf{SRCC}  & \textbf{LCC}   & \textbf{RMSE} & \textbf{OR}    & \textbf{SRCC}  & \textbf{LCC}   & \textbf{RMSE}  & \textbf{OR}   & \textbf{SRCC}  & \textbf{LCC}   & \textbf{RMSE} & \textbf{OR}\\ \hline

		\multirow{10}{*}{\textbf{2D FR}}         & \textbf{PSNR}          & 0.6026          & 0.6189          & 0.8031    &0.0045       & 0.8078          & 0.7830          & 1.2697  &0.0060   &0.7045 	&0.7035 	 	&1.5330 	&0.0195\\

		& \textbf{SSIM \cite{wang2004image}}          & 0.7346          & 0.7596          & 0.6650     &0.0000     & 0.7027          & 0.7123          & 1.4327 &0.0060 &0.6862 	&0.7455 	 	&1.4378 	&0.0156\\

		& \textbf{MS-SSIM \cite{wang2003multiscale}}         & 0.8266          & 0.8388          & 0.5566    &0.0000     & 0.7675          &0.7518           &1.3461  &0.0060 &0.6906 	&0.7539  	&1.4171 	&0.0117\\

		& \textbf{FSIM \cite{zhang2011fsim}}          & 0.8233          & 0.8318          & 0.5675    &0.0045     &0.7776 &0.7679 &1.3075   &0.0030   &0.7811 	&0.8139 	 	&1.2533 	&0.0039\\

		& \textbf{IWSSIM \cite{wang2011information}}        & 0.8352          & 0.8435          & 0.5492   &0.0000   &0.8124 &0.7966 &1.2340 &0.0030     &0.7111 	&0.7971 	 	&1.3024 	&0.0000 \\

		& \textbf{IFC \cite{sheikh2005information}}           & 0.5028          & 0.5393          & 0.8611  &0.0000   &0.7573 &0.7445 &1.3629   &0.0030    &0.4827 	&0.5946 	 	&1.7343 	&0.0156\\

		& \textbf{VIF \cite{sheikh2006image}}           & 0.6665          & 0.7032          & 0.7270   &0.0000  &0.7843 &0.7861 &1.2618 &0.0030     &0.0684 	&0.2533  	&2.0867 	&0.0469   \\

		& \textbf{NQM \cite{damera2000image}}           & 0.6508          & 0.6940          & 0.7362  &0.0045   &0.7202 &0.7361 &1.3817  &0.0060   &0.4601 	&0.5305 	&1.8285 	&0.0234   \\

		& \textbf{VSNR \cite{chandler2007vsnr}}          & 0.3961          & 0.5050          & 0.8826  &0.0182  &0.7427 &0.5787 &1.6651 &0.0179  &0.5542 	&0.6289  	&1.6770 	&0.0156\\

		& \textbf{HDR-VDP2 \cite{mantiuk2011hdr}}  &0.5555 &0.6300 &0.7941 &0.0045 &0.8608 &0.8385 &1.1123 &0.0000 &0.1888 	&0.3347 	&2.0327 	&0.0625 \\ \hline

		\multirow{3}{*}{\textbf{2D NR}}          & \textbf{BRISQUE \cite{mittal2012no}}       & 0.6687          & 0.7510          & 0.5619  &0.0000   &0.6724 &0.7597 &1.1317 &0.0000    &0.8239 &0.8843 &0.8325 &0.0000 \\

		& \textbf{NIQE \cite{mittal2012making}}          & 0.2086          & 0.2645          & 0.9861   &0.0045  &0.0665 &0.1950 &2.0022   &0.0327   &0.1386 	&0.1114  	&2.1436 	&0.0547 \\

		& \textbf{FRIQUEE \cite{ghadiyaram2017perceptual}}         & 0.6328          & 0.7213          & 0.5767  &0.0000    &0.6454 &0.7451 &1.1036 &0.0000  &0.7269 &0.8345 &0.9742 &0.0000   \\ \hline

		\textbf{3D FR}                           & \textbf{Chen \cite{chen2013full}}          & 0.5269          & 0.6070          & 0.8126  &0.0091   &0.7668 &0.7585 &1.3303  &0.0030   &0.6798 	&0.7722 	&1.3706 	&0.0078  \\ \hline

		\multirow{2}{*}{\textbf{3D NR}}          & \textbf{SINQ \cite{liu2017binocular}}          & 0.8029          & 0.8362          & 0.5124  &0.0000    &0.8524 &0.8612 &0.9939 &0.0000    &0.8682 &0.8968 &0.9653 &0.0000\\

		& \textbf{BSVQE \cite{chen2018blind}}         & 0.8179          & 0.8425          & 0.4801  &0.0000     &0.8570 &0.8751 &0.9561 &0.0000  &0.8449 &0.8992 &0.8514 &0.0000  \\ \hline

		\multirow{5}{*}{\textbf{Multi-view FR}} & \textbf{MP-PSNR Full \cite{sandic2015dibr}}  & 0.5335          & 0.4766          & 0.8989   &0.0000      &0.7203 &0.6730 &1.5099     &0.0089  &0.8449 &0.8992 &0.8514 &0.0000   \\

		& \textbf{MP-PSNR Reduc \cite{sandic2016multi}} & 0.5374          & 0.4765          & 0.8989   &0.0000    &0.7210 &0.6747 &1.5067  &0.0089   &0.6716 	&0.6926 	&1.5559 	&0.0117  \\

		& \textbf{MW-PSNR Full \cite{sandic2015dibr1}}  & 0.5147          & 0.4758          & 0.8993   &0.0000      &0.7232 &0.6770 &1.5023   &0.0089   &0.6620 	&0.6505 	&1.6382 	&0.0117 \\

		& \textbf{MW-PSNR Reduc \cite{sandic2015dibr1}} & 0.5326          & 0.4766          & 0.8989   &0.0000    &0.7217 &0.6757 &1.5048  &0.0089   &0.6769 	&0.6903 	&1.5607 	&0.0117  \\

		& \textbf{3DSwIM \cite{battisti2015objective}}        & 0.4320          & 0.5262          & 0.8695   &0.0182    &0.5565 &0.5489 &1.7063 &0.0119  &0.4053 	&0.4707 	&1.9032 	&0.0234 \\ \hline

		\textbf{Multi-view NR}                  & \textbf{APT \cite{gu2018model}}           & 0.3058          & 0.4087          & 0.9332  &0.0045   &0.0710 &0.0031 &2.0413 &0.0357  &0.5105   & 0.5249 &1.8361 &0.0234\\ \hline

		\textbf{LFI FR}                  & \textbf{Fang \cite{fang2018light}}           & -          & -          & -   &-   &0.7942      &0.8065     &1.2300  &-      & -          & -          & -   &-   \\ \hline

		\textbf{LFI RR}                  & \textbf{LF-IQM \cite{8632960}}           & 0.4478          & 0.5193          & 0.8584  &0.0227  &0.3334   &0.4360   &1.8038  &0.0149  &0.1701 &0.2101 &1.0279 &0.0392\  \\ \hline

		\textbf{LFI NR}                          & \textbf{Proposed NR-LFQA}           & \textbf{0.9032} & \textbf{0.9206} & \textbf{0.3876} & \textbf{0.0000} & \textbf{0.9119} & \textbf{0.9155} & \textbf{0.7556} & \textbf{0.0000} &\textbf{0.8803} &\textbf{0.9105} &\textbf{0.8300} &\textbf{0.0000} \\ \hline

	\end{tabular}

\end{table*}

\renewcommand\arraystretch{1.3}
\begin{table*}[!htb]
	\centering
	\scriptsize
	\caption{Performance Comparison on VALID Dataset.}
	\begin{tabular}{c|c|cccc|cccc}
		\hline
		\multicolumn{1}{c|}{} & \multicolumn{1}{c|}{} & \multicolumn{4}{c|}{\textbf{VALID-8bit }}  & \multicolumn{4}{c}{\textbf{VALID-10bit }}\\ \hline
		\textbf{Type}                            & \textbf{Metrics}       & \textbf{SRCC}  & \textbf{LCC}   & \textbf{RMSE} & \textbf{OR}    & \textbf{SRCC}  & \textbf{LCC}   & \textbf{RMSE}  & \textbf{OR}  \\ \hline
		\multirow{10}{*}{\textbf{2D FR}}         & \textbf{PSNR}         &0.9620 &0.9681 &0.3352 &0.0000  &0.9467  &0.9524  &0.2935 &0.0000\\
		& \textbf{SSIM \cite{wang2004image}}           & 0.9576          & 0.9573          & 0.3868 &0.0000  &0.9326 &0.9375 &0.3348 &0.0000\\
		& \textbf{MS-SSIM \cite{wang2003multiscale}}         &0.9593 &0.9658   &0.3473 &0.0000  &0.9432 &0.9484 &0.3051 &0.0000\\
		& \textbf{FSIM \cite{zhang2011fsim}}         & 0.9695    &0.9798         &0.2678  &0.0000 &- &- &- &-\\
		& \textbf{IWSSIM \cite{wang2011information}}        &0.9674 &0.9764 &0.2892   &0.0000 &\textbf{0.9499} &\textbf{0.9617} &\textbf{0.2638} &0.0000 \\
		& \textbf{IFC \cite{sheikh2005information}}            &0.9693 &\textbf{0.9909} &\textbf{0.1800}  &0.0000   &- &- &- &- \\
		& \textbf{VIF \cite{sheikh2006image}}             &\textbf{0.9749} &0.9870 &0.2150    &0.0000 &-  &-  &- &-  \\
		& \textbf{NQM \cite{damera2000image}}              &0.9055 &0.9194 &0.5266   &0.0000  &0.8410 &0.8582 &0.4940 &0.0000\\
		& \textbf{VSNR \cite{chandler2007vsnr}}             &0.9359 &0.9324 &0.4838   &0.0000  &- &- &- &- \\
		& \textbf{HDR-VDP2 \cite{mantiuk2011hdr}}  &0.9623 &0.9785 &0.2758 &0.0000 &0.9371 &0.9528 &0.2921 &0.0000 \\ \hline
		\multirow{3}{*}{\textbf{2D NR}}          & \textbf{BRISQUE \cite{mittal2012no}}       &0.9222 &0.9849 &0.2017   &0.0000 &0.9027 &0.9347 &0.2838 &0.0000 \\
		& \textbf{NIQE \cite{mittal2012making}}          &0.8636 &0.9524 &0.4080  &0.0000  &- &- &- &- \\
		& \textbf{FRIQUEE \cite{ghadiyaram2017perceptual}}        &0.9157 &0.9836 &0.2160   &0.0000 &0.8559 &0.8986 &0.3497 &0.0000  \\ \hline
		\textbf{3D FR}                           & \textbf{Chen \cite{chen2013full}}           &0.9642 &0.9738 &0.3046   &0.0000 &- &- &- &-\\ \hline
		\multirow{2}{*}{\textbf{3D NR}}          & \textbf{SINQ \cite{liu2017binocular}}         &0.9222 &0.9849 &0.2070    &0.0000  &0.9021 &0.9348 &0.2722 &0.0000 \\
		& \textbf{BSVQE \cite{chen2018blind}}         &0.9222 &0.9814 &0.2180    &0.0000  &- &- &- &-\\ \hline
		\multirow{5}{*}{\textbf{Multi-view FR}} & \textbf{MP-PSNR Full \cite{sandic2015dibr}}   &0.9730 &0.9852 &0.2291  &0.0000  &0.3830 &0.3582 &0.8986 &0.0000\\
		& \textbf{MP-PSNR Reduc \cite{sandic2016multi}}  &0.9744 &0.9859 &0.2237   &0.0000 & 0.3826 &0.3506 &0.9013 &0.0000\\
		& \textbf{MW-PSNR Full \cite{sandic2015dibr1}}     &0.9597 &0.9677 &0.3376   &0.0000  &0.3764 &0.3556 &0.8995 &0.0000\\
		& \textbf{MW-PSNR Reduc \cite{sandic2015dibr1}} &0.9648 &0.9751 &0.2970   &0.0000  &0.3815 &0.3563 &0.8993 &0.0100\\
		& \textbf{3DSwIM \cite{battisti2015objective}}         &0.7950 &0.7876 &0.8248  &0.0000 &0.7869 &0.7401 &0.6472 &0.0000\\ \hline
		\textbf{Multi-view NR}                  & \textbf{APT \cite{gu2018model}}       &0.4699 &0.6452 &1.0228   &0.0000  &- &- &- &-\\ \hline
		\textbf{LFI RR}                  & \textbf{LF-IQM \cite{8632960}}             &0.3571   &0.4708   &1.0712    &0.0000 &0.3692 &0.4828 &0.8137 &0.0000 \\ \hline
		\textbf{LFI NR}                          & \textbf{Proposed NR-LFQA}            &\textbf{0.9286}   &\textbf{0.9799}   &\textbf{0.2380}  &\textbf{0.0000} &\textbf{0.9228} &\textbf{0.9517} &\textbf{0.2769} &\textbf{0.0000}\\ \hline
	\end{tabular}
\end{table*}

Since the pixels in each row of the EPI come from different SAIs, the relationship between the pixels of each row can reflect the relationship of the SAIs. This property indicates that we can analyze the relative relationship of pixels in EPI to measure the change in angular consistency. In addtion, measuring the relationship between pixels at different distances can capture local angular consistency information.

\begin{figure*}[!htb]
\centering
\begin{minipage}{0.19\linewidth}
  \centerline{\includegraphics[width=3.9cm]{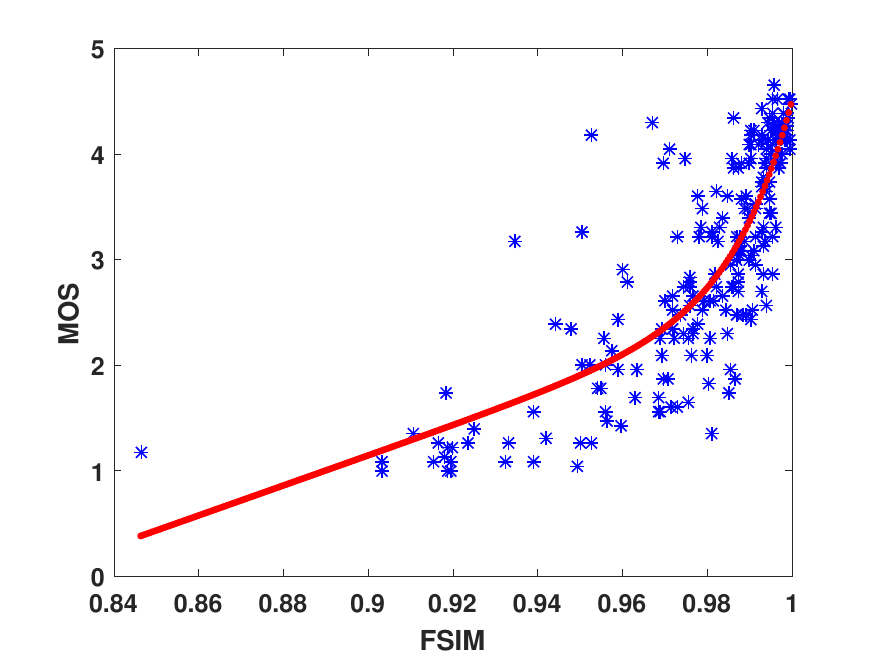}}
  \centerline{(a)}
\end{minipage}
%\hfill
\begin{minipage}{0.19\linewidth}
  \centerline{\includegraphics[width=3.9cm]{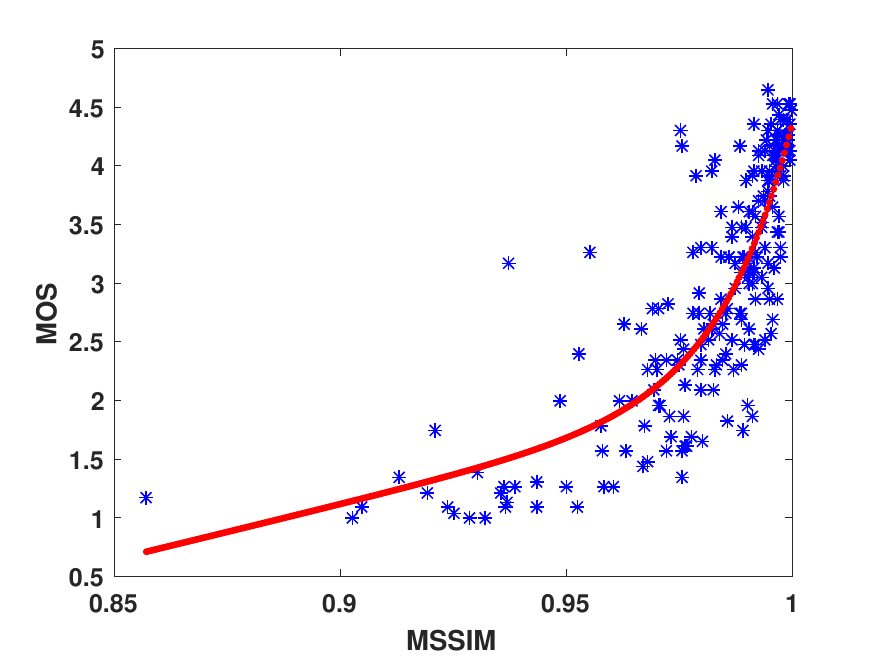}}
  \centerline{(b)}
\end{minipage}
%\hfill
\begin{minipage}{0.19\linewidth}
  \centerline{\includegraphics[width=3.9cm]{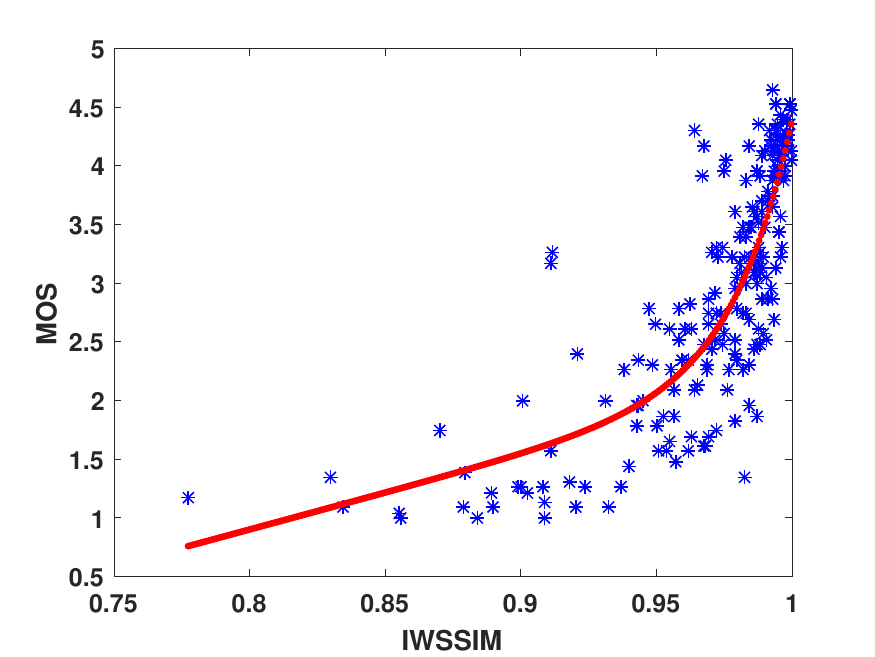}}
  \centerline{(c)}
\end{minipage}
%\hfill
\begin{minipage}{0.19\linewidth}
  \centerline{\includegraphics[width=3.9cm]{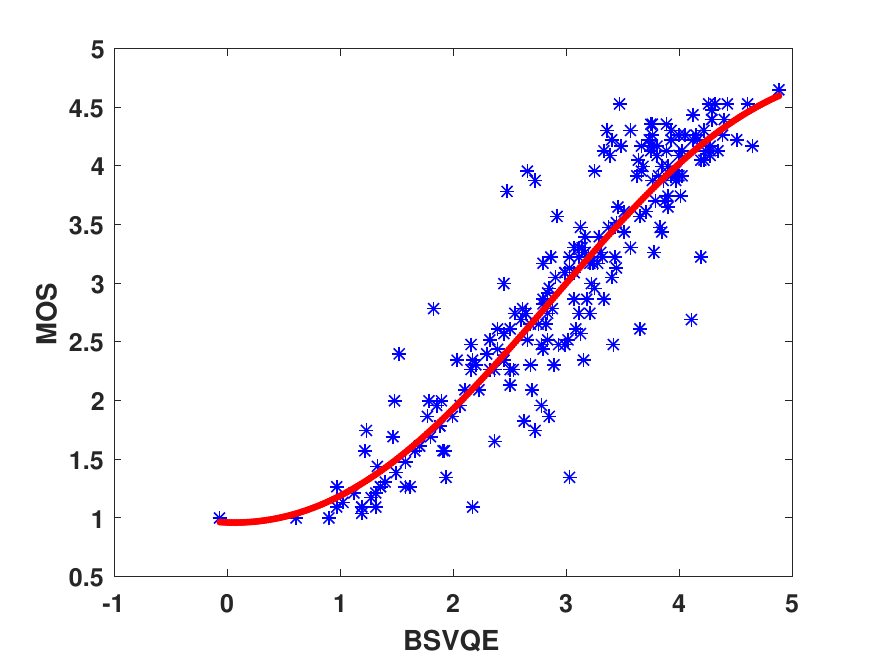}}
  \centerline{(d)}
\end{minipage}
%\hfill
\begin{minipage}{0.19\linewidth}
  \centerline{\includegraphics[width=3.9cm]{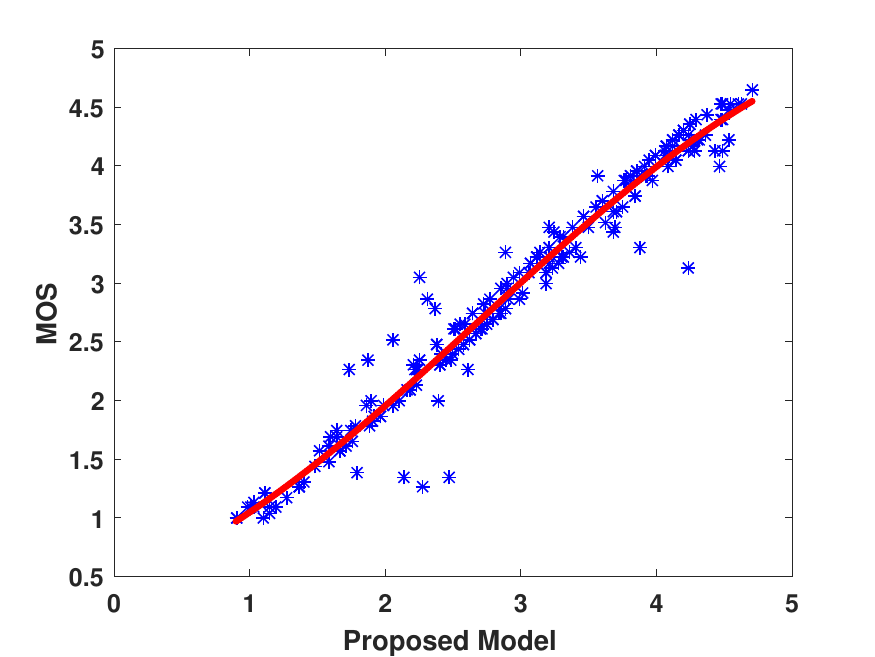}}
  \centerline{(e)}
\end{minipage}
%\hfill
\caption{The scatter plots of predicted quality scores by different methods against the MOS values on the Win5-LID. The horizontal axis in each figure denotes predicted quality scores and the vertical axis in each figure represents the MOS values. The red line is the fitted curve. (a) FSIM; (b) MS-SSIM; (c) IWSSIM; (d) BSVQE; (e) Proposed.}
\centering
\end{figure*}

%\begin{figure*}[!htb]
%\centering
%\begin{minipage}{0.19\linewidth}
%  \centerline{\includegraphics[width=3.9cm]{fig/mpi-lfa_psnr_scatter.eps}}
%  \centerline{(a)}
%\end{minipage}
%%\hfill
%\begin{minipage}{0.19\linewidth}
%  \centerline{\includegraphics[width=3.9cm]{fig/mpi-lfa_iwssim_scatter.eps}}
%  \centerline{(b)}
%\end{minipage}
%%\hfill
%\begin{minipage}{0.19\linewidth}
%  \centerline{\includegraphics[width=3.9cm]{fig/mpi-lfa_sinq_scatter.eps}}
%  \centerline{(c)}
%\end{minipage}
%%\hfill
%\begin{minipage}{0.19\linewidth}
%  \centerline{\includegraphics[width=3.9cm]{fig/mpi-lfa_bsvqe_scatter.eps}}
%  \centerline{(d)}
%\end{minipage}
%%\hfill
%\begin{minipage}{0.19\linewidth}
%  \centerline{\includegraphics[width=3.9cm]{fig/mpi-lfa_nlfa_scatter.eps}}
%  \centerline{(e)}
%\end{minipage}
%%\hfill
%\caption{The scatter plots of predicted quality scores by different methods against the MOS values on the MPI-LFA. The horizontal axis in each %figure denotes predicted quality scores and the vertical axis in each figure represents the MOS values. The red line is the fitted curve. (a) %PSNR; (b) IWSSIM; (c) SINQ; (d) BSVQE; (e) Proposed.}
%\centering
%\end{figure*}

As a result, we propose Weighted Local Binary Pattern (WLBP) to capture the relationship between different SAIs.
The LBP is able to extract local distribution information, which has proven its effectiveness in many works and has shown good performance for the evaluation of 2D-IQA tasks \cite{ojala2002multiresolution,satpathy2014lbp,nanni2012survey,freitas2016blind,zhang2015blind,zhang2013no}.
We first calculate the local rotation invariant uniform LBP operator $L_{u^*,s^*}^{R,P}$ of the vertical EPI $E_{u^*,s^*}$ by:

\begin{footnotesize}
\begin{equation}
L_{u^*,s^*}^{R,P}(E_{u^*,s^*}^c)= \left\{
\begin{aligned}
\sum_{p=0}^{P-1}\theta(E_{u^*,s^*}^p-E_{u^*,s^*}^c) & & \psi(\hat{L}_{u^*,s^*}^{R,P}) \leq 2 \\
P+1  & & otherwise,
\end{aligned}
\right.
\end{equation}
\end{footnotesize}
\noindent{where} $R$ is the radius value and $P$ indicates the number of neighboring points.
$E_{u^*,s^*}^c$ represents a center pixel at the position $(x_c,y_c)$ in the corresponding EPIs and $E_{u^*,s^*}^p$ is a neighboring pixel $(x_p,y_p)$ surrounding $E_{u^*,s^*}^c$:
\begin{equation}
x_p=x_c+R\cos(2\pi\frac{p}{P}) \quad and \quad y_p=y_c-R\sin(2\pi\frac{p}{P}),
\end{equation}
where $p\in\{1,2...P\}$ is the number of neighboring pixels sampled by a distance $R$ from $E_{u^*,s^*}^c$ to $E_{u^*,s^*}^p$.
In this case, $\theta(z)$ is the step function and defined by:
\begin{equation}
\theta(z)=\left\{
\begin{aligned}
1 & & z \geqslant T \\
0 & & otherwise,
\end{aligned}
\right.
\end{equation}
where $T$ indicates the threshold value. In addition, $\psi$ is used to compute the number of bitwise transitions:

\begin{small}
\begin{equation}
\begin{aligned}
%\resizebox{.85\hsize}{!}
\psi(L_{u^*,s^*}^{R,P})= ||\theta(E_{u^*,s^*}^{P-1}-E_{u^*,s^*}^c)-\theta(E_{u^*,s^*}^0-E_{u^*,s^*}^c)|| \\
+\sum_{p=1}^{P-1}||\theta(E_{u^*,s^*}^p-E_{u^*,s^*}^c)-\theta(E_{u^*,s^*}^{p-1}-E_{u^*,s^*}^c)||,
%\Delta(E_{P-1},E_0)+\sum_{p=1}^{P-1}\Delta(E_p,E_p-1)
\end{aligned}
\end{equation}
\end{small}
and $\hat{L}_{u^*,s^*}^{R,P}$ is rotation-invariant operator:

\begin{footnotesize}
\begin{equation}
%\begin{footnotesize}
%\resizebox{1\hsize}{!}{$\hat{L}_{u^*,s^*}^{R,P}(E_{u^*,s^*}^c)=min\{ROR(\sum\limits_{p=0}^{P-1}\theta(E_{u^*,s^*}^p-E_{u^*,s^*}^c)2^p,k)\}$}
\hat{L}_{u^*,s^*}^{R,P}(E_{u^*,s^*}^c)=min\{ROR(\sum\limits_{p=0}^{P-1}\theta(E_{u^*,s^*}^p-E_{u^*,s^*}^c)2^p,k)\},
%\end{footnotesize}
\end{equation}
\end{footnotesize}
\noindent{where} $k\in\{0,1,2...,P-1\}$ and $ROR(\beta,k)$ is the circular bit-wise right shift operator that shifts the tuple $\beta$ by $k$ positions. Finally, we obtain $L_{u^*,s^*}^{R,P}$ with a length of $P+2$.

\begin{figure*}[!htb]
\centering
\begin{minipage}{0.47\linewidth}
    \centerline{\includegraphics[width=9cm]{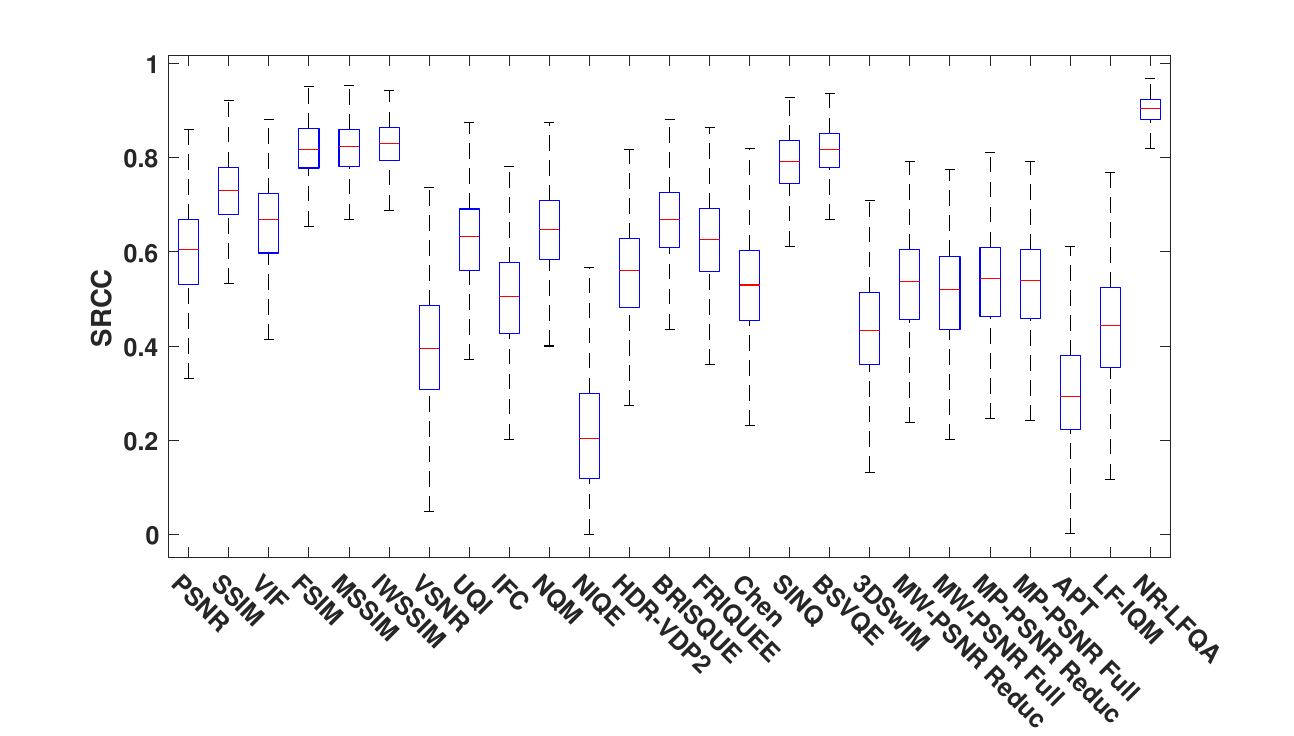}}
    \centerline{(a)}
\end{minipage}
%\vfill
\begin{minipage}{0.47\linewidth}
    \centerline{\includegraphics[width=9cm]{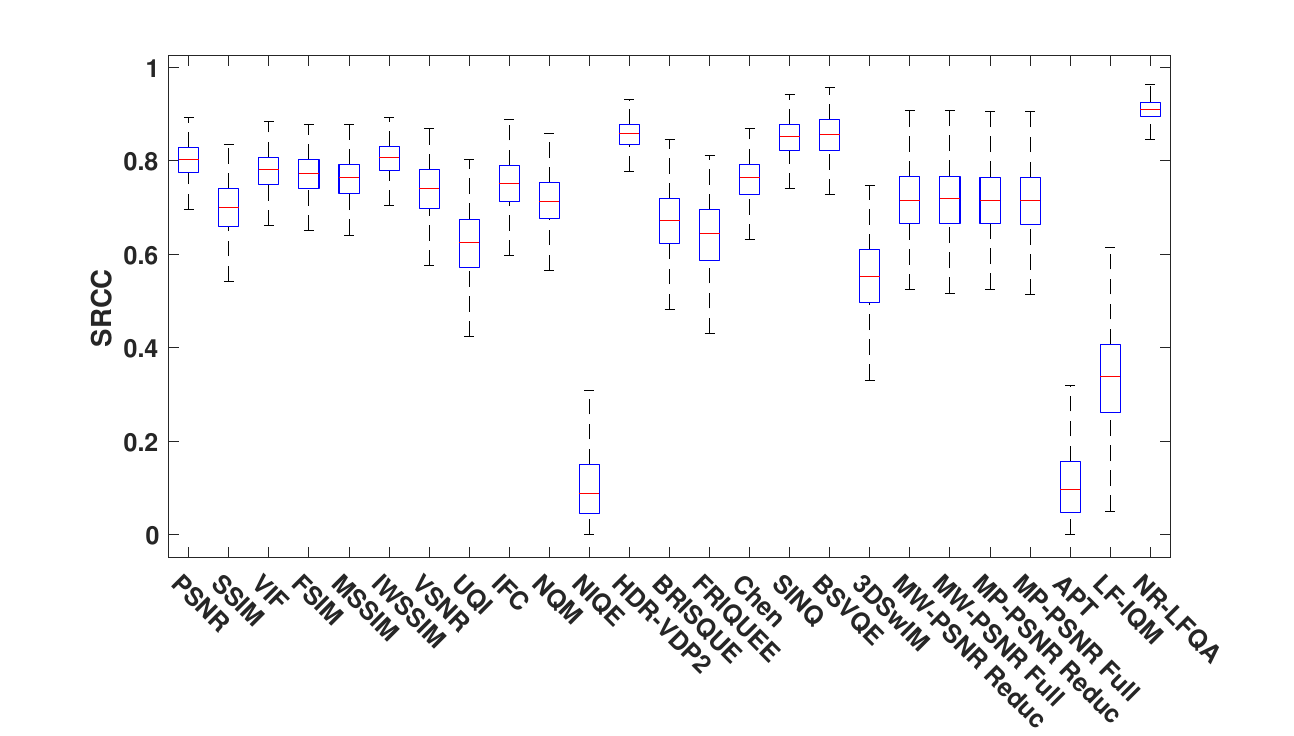}}
    \centerline{(b)}
\end{minipage}
\caption{Box plot of SRCC distributions of algorithms over 1000 trials. (a) On Win5-LID dataset; (b) On MPI-LFA dataset.}
\centering
\end{figure*}

For a LFI, there exist many EPIs in the vertical and horizontal directions. If we concatenate the LBP features that extracted from each EPI, it will induce a dimensional disaster.
To reduce the feature dimension, entropy weighting is adopted because some EPIs have little information and their LBP features are mainly concentrated in one statistical direction. This represents that these EPIs contain less angular consistency information and their entropy values are close to zero.
Therefore, we can obtain the WLBP of vertical EPI $E_{u^*,s^*}$ as follows:
\begin{equation}
Lver_{R,P}=\frac{\sum\limits_{u=1}^{U}\sum\limits_{s=1}^{S}w_{u,s}^{R,P}.*L_{u,s}^{R,P}}{\sum\limits_{u=1}^{U}\sum\limits_{s=1}^{S}w_{u,s}^{R,P}},
\end{equation}
where the entropy of $L_{u,s}^{R,P}$ is computed as weight $w_{u,s}^{R,P}$. At the same time, we adopt the same operation to obtain the WLBP features $Lhor_{R,P}$ of the horizontal EPI $E_{v^*,t^*}$.

In our implement, we set $R\in\{1,2,3\}$, $P=3 \times R$ and $T=R/2$. Finally, we combine all features to obtain feature $F_{WLBP}$ as follows:
\begin{equation}
F_{WLBP}=\{Lver_{R,P},Lhor_{R,P}\}.
\end{equation}

Fig. 8 illustrates the WLBP histogram of various distortion types with $R=1$ and $P=8$. The LFIs are selected from MPI-LFA dataset \cite{adhikarla2017towards}. Specifically, the WLBP features are extracted from horizontal and vertical EPIs separately. For MPI-LFA dataset, the LFIs only contain horizontal EPIs. Moreover, for the WLBP feature with parameter $(R, P)$, we can generate $P+2$ histogram bins \cite{ojala2002multiresolution}. Therefore, when $P=8$, the corresponding feature dimension is $P+2=10$. Furthermore, the histogram of WLBP for NN distortion with different distortion levels for $R=1$ and $P=8$ is shown in Fig. 9.  As the figures show that there is a significant difference in the WLBP distribution for different distortion types, and the WLBP distribution changes significantly as the distortion level increases. Overall, WLBP can effectively capture the characteristics of local angular consistency degradation in EPI.

\subsection{Quality Evaluation}
In order to conduct the quality assessment, we train a regression model to map the final feature vector $F_{Final}=\{F_{LCN},F_{GDD},F_{WLBP}\}$ space to quality scores. In our implementation, we utilize support vector regression (SVR) model, which can automatically learn the weights of these features from the data distribution. In other words, the weights of these proposed features are different. Moreover, the SVR model has been applied to many image quality assessment problems \cite{zhou20163d,chen2018blind,liu2017binocular} and has demonstrated its effectiveness The LIBSVM package \cite{chang2011libsvm} is exploited to implement the SVR with a radial basis function (RBF) kernel.

\renewcommand\arraystretch{1.3}
\begin{table*}[]
\tiny
\centering
\caption{Performance Comparison Between Row and Column Methods With T-test on Win5-LID Dataset. The Symbol `1', `0', or `-1' Represents that the Row Method is Statistically Better, Indistinguishable, or Worse Than The Column Algorithm. Due to Space Constraints, Here We Use the Referenced Number Instead of The Algorithm Name.}
\begin{tabular}{|p{0.6cm}<{\centering}|p{0.4cm}<{\centering}|c|c|c|c|c|c|c|c|c|c|c|c|c|c|c|c|c|c|c|c|c|c|c|}
\hline
\multicolumn{1}{|c|}{\textbf{}} & PSNR & \cite{wang2004image} & \cite{sheikh2006image} & \cite{zhang2011fsim} & \cite{wang2003multiscale} & \cite{wang2011information} & \cite{chandler2007vsnr} & \cite{sheikh2005information} & \cite{damera2000image} & \cite{mittal2012making} & \cite{mittal2012no} & \cite{ghadiyaram2017perceptual} & \cite{chen2013full} & \cite{liu2017binocular} & \cite{chen2018blind} & \cite{battisti2015objective} & \cite{sandic2015dibr1} & \cite{sandic2015dibr1} & \cite{sandic2016multi} & \cite{sandic2015dibr}  & \cite{gu2018model} &\cite{8632960} & \textbf{Proposed} \\ \hline
PSNR          & 0      & -1     & -1    & -1     & -1      & -1       & 1      & 1     & -1    & 1      & -1        & -1      & 1      & -1     & -1      & 1        & 1               & 1              & 1               & 1              & 1     & 1      & -1       \\ \hline
\cite{wang2004image}          & 1      & 0      & 1     & -1     & -1      & -1       & 1      & 1     & 1     & 1      & 1         & 1       & 1      & -1     & -1      & 1        & 1               & 1              & 1               & 1              & 1     & 1      & -1       \\ \hline
\cite{sheikh2006image}           & 1      & -1     & 0     & -1     & -1      & -1       & 1      & 1     & 1     & 1      & 0         & 1       & 1      & -1     & -1      & 1        & 1               & 1              & 1               & 1              & 1     & 1      & -1       \\ \hline
\cite{zhang2011fsim}          & 1      & 1      & 1     & 0      & 0       & -1       & 1      & 1     & 1     & 1      & 1         & 1       & 1      & 1      & 0       & 1        & 1               & 1              & 1               & 1              & 1     & 1      & -1       \\ \hline
\cite{wang2003multiscale}         & 1      & 1      & 1     & 0      & 0       & -1       & 1      & 1     & 1     & 1      & 1         & 1       & 1      & 1      & 1       & 1        & 1               & 1              & 1               & 1              & 1     & 1      & -1       \\ \hline
\cite{wang2011information}         & 1      & 1      & 1     & 1      & 1       & 0        & 1      & 1     & 1     & 1      & 1         & 1       & 1      & 1      & 1       & 1        & 1               & 1              & 1               & 1              & 1     & 1      & -1       \\ \hline
\cite{chandler2007vsnr}          & -1     & -1     & -1    & -1     & -1      & -1       & 0      & -1    & -1    & 1      & -1        & -1      & -1     & -1     & -1      & -1       & -1              & -1             & -1              & -1             & 1     & -1     & -1       \\ \hline
\cite{sheikh2005information}           & -1     & -1     & -1    & -1     & -1      & -1       & 1      & 0     & -1    & 1      & -1        & -1      & -1     & -1     & -1      & 1        & -1              & -1             & -1              & -1             & 1     & 1      & -1       \\ \hline
\cite{damera2000image}           & 1      & -1     & -1    & -1     & -1      & -1       & 1      & 1     & 0     & 1      & -1        & 1       & 1      & -1     & -1      & 1        & 1               & 1              & 1               & 1              & 1     & 1      & -1       \\ \hline
\cite{mittal2012making}          & -1     & -1     & -1    & -1     & -1      & -1       & -1     & -1    & -1    & 0      & -1        & -1      & -1     & -1     & -1      & -1       & -1              & -1             & -1              & -1             & -1    & -1     & -1       \\ \hline
\cite{mittal2012no}       & 1      & -1     & 0     & -1     & -1      & -1       & 1      & 1     & 1     & 1      & 0         & 1       & 1      & -1     & -1      & 1        & 1               & 1              & 1               & 1              & 1     & 1      & -1       \\ \hline
\cite{ghadiyaram2017perceptual}         & 1      & -1     & -1    & -1     & -1      & -1       & 1      & 1     & -1    & 1      & -1        & 0       & 1      & -1     & -1      & 1        & 1               & 1              & 1               & 1              & 1     & 1      & -1       \\ \hline
\cite{chen2013full}          & -1     & -1     & -1    & -1     & -1      & -1       & 1      & 1     & -1    & 1      & -1        & -1      & 0      & -1     & -1      & 1        & 0               & 1              & -1              & 0              & 1     & 1      & -1       \\ \hline
\cite{battisti2015objective}          & 1      & 1      & 1     & -1     & -1      & -1       & 1      & 1     & 1     & 1      & 1         & 1       & 1      & 0      & -1      & 1        & 1               & 1              & 1               & 1              & 1     & 1      & -1       \\ \hline
\cite{chen2018blind}         & 1      & 1      & 1     & 0      & -1      & -1       & 1      & 1     & 1     & 1      & 1         & 1       & 1      & 1      & 0       & 1        & 1               & 1              & 1               & 1              & 1     & 1      & -1       \\ \hline
\cite{battisti2015objective}        & -1     & -1     & -1    & -1     & -1      & -1       & 1      & -1    & -1    & 1      & -1        & -1      & -1     & -1     & -1      & 0        & -1              & -1             & -1              & -1             & 1     & -1     & -1       \\ \hline
\cite{sandic2015dibr1} & -1     & -1     & -1    & -1     & -1      & -1       & 1      & 1     & -1    & 1      & -1        & -1      & 0      & -1     & -1      & 1        & 0               & 1              & 0               & 0              & 1     & 1      & -1       \\ \hline
\cite{sandic2015dibr1}  & -1     & -1     & -1    & -1     & -1      & -1       & 1      & 1     & -1    & 1      & -1        & -1      & -1     & -1     & -1      & 1        & -1              & 0              & -1              & -1             & 1     & 1      & -1       \\ \hline
\cite{sandic2016multi}  & -1     & -1     & -1    & -1     & -1      & -1       & 1      & 1     & -1    & 1      & -1        & -1      & 1      & -1     & -1      & 1        & 0               & 1              & 0               & 0              & 1     & 1      & -1       \\ \hline
\cite{sandic2015dibr}  & -1     & -1     & -1    & -1     & -1      & -1       & 1      & 1     & -1    & 1      & -1        & -1      & 0      & -1     & -1      & 1        & 0               & 1              & 0               & 0              & 1     & 1      & -1       \\ \hline
\cite{gu2018model}           & -1     & -1     & -1    & -1     & -1      & -1       & -1     & -1    & -1    & 1      & -1        & -1      & -1     & -1     & -1      & -1       & -1              & -1             & -1              & -1             & 0     & -1     & -1       \\ \hline
\cite{8632960}          & -1     & -1     & -1    & -1     & -1      & -1       & 1      & -1    & -1    & 1      & -1        & -1      & -1     & -1     & -1      & 1        & -1              & -1             & -1              & -1             & 1     & 0      & -1       \\ \hline
\textbf{Proposed}        & 1      & 1      & 1     & 1      & 1       & 1        & 1      & 1     & 1     & 1      & 1         & 1       & 1      & 1      & 1       & 1        & 1               & 1              & 1               & 1              & 1     & 1      & 0        \\ \hline
\end{tabular}
\end{table*}

We select four evaluation criteria to measure the performance of objective quality assessment models, namely Spearman rank-order correlation coefficient (SRCC), linear correlation coefficient (LCC), root-mean-square-error (RMSE) and outlier ratio (OR). The SRCC measures the monotonicity, while the LCC evaluates the linear relationship between predict scores and MOS values. The RMSE and OR measure the prediction accuracy and prediction consistency, respectively. The SRCC and LCC values closing to 1 represent high positive correlation and a lower RMSE/OR value indicates a better performance. Additionally, each dataset is randomly divided into 80\% for training and 20\% for testing. We perform 1000 iterations of cross validation on each dataset, and provide the median SRCC, LCC, RMSE and OR performance as the final measurement.
Before computing the LCC, RMSE and OR, a non-linear function is employed by:
\begin{equation}
q'=\beta_1\{\frac{1}{2}-\frac{1}{1+exp[\beta_2(q-\beta_3)]}\}+\beta_4q+\beta_5,
\end{equation}
where $q$ is the output of a method. The parameters $\beta_{1 \cdots  5}$  are optimized to minimize a given goodness-of-fit measure. Although the prediction range of the model is different, all prediction scores can be guaranteed to be evaluated on the same scale after logistic function fitting.

\renewcommand\arraystretch{1.3}
\begin{table}[!htb]
\centering
\caption{Performance Comparison of Existing Algorithms on EPIs.}
\begin{tabular}{p{1.9cm}<{\centering}|p{0.6cm}<{\centering} p{0.6cm}<{\centering} p{0.7cm}<{\centering}|p{0.6cm} p{0.6cm} p{0.6cm}}
\hline
& \multicolumn{3}{c|}{\textbf{Win5-LID}}       & \multicolumn{3}{c}{\textbf{MPI-LFA}}   \\ \hline
\textbf{Metrics} & \textbf{SRCC}  & \textbf{LCC}    & \textbf{RMSE}  & \textbf{SRCC}  & \textbf{LCC}    & \textbf{RMSE} \\ \hline
\textbf{PSNR}    & 0.7190          & 0.7213              & 0.7083       & 0.7198          & 0.6972                   &    1.4634     \\
\textbf{FSIM \cite{zhang2011fsim}}    & 0.7545          & 0.7780          & 0.6424      & 0.7920          & 0.7818          & 1.2730    \\
\textbf{IFC \cite{sheikh2005information}}     & 0.5323          & 0.5010          & 0.8850        & 0.7265          & 0.7394          & 1.3743  \\          \hline
\textbf{BRISQUE \cite{mittal2012no}} & 0.7727          & 0.8473                   & 0.4890        & 0.7864          & 0.7850                    & 1.0453  \\
\textbf{FRIQUEE \cite{ghadiyaram2017perceptual}}   & 0.6225          & 0.6648                   & 0.5784       & 0.7523          & 0.7652                   & 1.0753   \\ \hline
\textbf{Fang \cite{fang2018light}}           & -          & -          & -      &0.7942      &0.8065     &1.2300       \\ \hline
\textbf{Proposed}     & \textbf{0.8829} & \textbf{0.9015}  & \textbf{0.3993} & \textbf{0.8905} & \textbf{0.8954}  & \textbf{0.8259} \\ \hline
\end{tabular}
\end{table}

\renewcommand\arraystretch{1.3}
\begin{table}[!htb]
\centering
\caption{Performance of Three Quality Components on Win5-LID and MPI-LFA Datasets.}
\begin{tabular}{p{1.2cm}<{\centering}|p{0.8cm}<{\centering} p{0.8cm}<{\centering} p{0.7cm}<{\centering}|p{0.8cm} p{0.8cm} p{0.6cm}}
%\begin{tabular}{c|ccc|ccc}
\hline
& \multicolumn{3}{c|}{\textbf{Win5-LID}}       & \multicolumn{3}{c}{\textbf{MPI-LFA}}   \\ \hline
\textbf{Features} & \textbf{SRCC}  & \textbf{LCC}    & \textbf{RMSE}  & \textbf{SRCC}  & \textbf{LCC}    & \textbf{RMSE} \\ \hline
\textbf{$F_{LCN}$} & 0.7464          & 0.8370       & 0.5128       &0.5806     &0.6260     &1.2891 \\
\textbf{$F_{GDD}$}  & 0.7431         & 0.7745        & 0.6090       &0.5621 &0.6205 &1.1383          \\
\textbf{$F_{WLBP}$} & 0.8528         & 0.8771        & 0.4006       &0.8437 &0.8573 &0.8900          \\ \hline
\end{tabular}
\end{table}

For OR computation, according to \cite{video2000final,ma2018joint,li2018has}, we calculate the standard deviation of the testing set. If the difference between the predicted score and the subjective score is more than 2 times the standard deviation, this can be defined as the outlier. Specifically, the OR represents the mapped objective scores deviating from the subjective ratings in 2 standard deviation as:
\begin{equation}
OR=\frac{\sum\limits_{i=1}^{N}{(|{{s}_{i}}-{{f}_{i}}|>2{{\sigma }_{i}})}}{N},
\end{equation}
where $N$ is the size of the testing set. ${s}_{i}$ and ${f}_{i}$ denote the $i$-th subjective score and the $i$-th mapped objective score after the non-linear mapping, respectively. ${\sigma }_{i}$ is the $i$-th standard deviation of the subjective scores.

\subsection{Comparison with Other Objective Metrics}
\renewcommand\arraystretch{1.3}
\begin{table*}[!htb]
\scriptsize
\centering
\caption{SRCC of Different Distortion Types on Win5-LID and MPI-LFA Datasets.}
\begin{tabular}{c|c|cccc|cccccc}
\hline
& & \multicolumn{4}{c|}{\textbf{Win5-LID}}       & \multicolumn{6}{c}{\textbf{MPI-LFA}}   \\ \hline
\textbf{Type} & \textbf{Metrics} & \textbf{HEVC}   & \textbf{JPEG}   & \textbf{LN}     & \textbf{NN}   & \textbf{HEVC}   & \textbf{DQ}     & \textbf{OPT}    & \textbf{Linear} & \textbf{NN}     & \textbf{GAUSS}  \\ \hline
\multirow{10}{*}{\textbf{2D FR}} & \textbf{PSNR}    & 0.8183          & 0.8302          & 0.8135          & 0.8734        & 0.9226          & 0.7664          & 0.6842          & 0.9090          & 0.8895          & 0.9239  \\
& \textbf{SSIM \cite{wang2004image}}    & 0.9351          & 0.8390          & 0.8501          & 0.8226        & 0.9769          & 0.4468          & 0.5609          & 0.8055          & 0.7682          & 0.9800  \\
& \textbf{MS-SSIM \cite{wang2003multiscale}}   & 0.9731          & 0.8947          & 0.8789          & 0.8346      & 0.9791          & 0.7087          & 0.6556          & 0.8763          & 0.8775          & 0.9702    \\
& \textbf{FSIM \cite{zhang2011fsim}}    & 0.9756          & 0.9006          & 0.8876          & 0.8661      & 0.9844          & 0.7631          & 0.6969          & 0.8745          & 0.8704          & 0.9502    \\
& \textbf{IWSSIM \cite{wang2011information}}  & 0.9690          & 0.9046          & 0.8851          & 0.8313       & 0.9831          & 0.7628          & 0.7336          & 0.9144          & 0.9256          & 0.9662   \\
& \textbf{IFC \cite{sheikh2005information}}     & 0.8634          & 0.7299          & 0.6179          & 0.8037     & 0.9430          & 0.8462          & 0.7909          & 0.9147          & 0.9261          & 0.8852     \\
& \textbf{VIF \cite{sheikh2006image}}     & 0.9627          & 0.8982          & 0.8287          & 0.8816      & 0.9818          & 0.7057          & 0.6445          & 0.9163          & 0.9048          & 0.9506    \\
& \textbf{NQM \cite{damera2000image}}     & 0.8861          & 0.7878          & 0.7086          & 0.7214     & 0.9515          & 0.6598          & 0.5596          & 0.8444          & 0.9264          & 0.9333      \\
& \textbf{VSNR \cite{chandler2007vsnr}}    & 0.9162          & 0.8861          & 0.0208          & 0.1705       & \textbf{0.9871}          & 0.6082          & 0.6634          & 0.8333          & 0.8256          & 0.9462   \\
& \textbf{HDR-VDP2 \cite{mantiuk2011hdr}}  &0.7322 &0.5282 &0.7912 &0.8711 &0.8065 &0.7448 &0.7953 &0.8632 &0.9008 &0.9309 \\ \hline
\multirow{3}{*}{\textbf{2D NR}} & \textbf{BRISQUE \cite{mittal2012no}} & 0.9152          & 0.9394          & 0.8268          & 0.2516     & 0.9429          & 0.6121          & 0.7059          & 0.7333          & 0.0941          & 0.8286     \\
& \textbf{NIQE \cite{mittal2012making}}    & 0.1332          & 0.2161          & 0.3749          & 0.1514      & 0.9186          & 0.0500          & 0.1226          & 0.2616          & 0.1457          & 0.1413      \\
& \textbf{FRIQUEE \cite{ghadiyaram2017perceptual}}   & 0.8545          & 0.9273          & 0.8303          & 0.2727     & 0.9429          & 0.7333          & 0.7912          & 0.6364          & 0.1059          & 0.9429     \\ \hline
\textbf{3D FR} & \textbf{Chen \cite{chen2013full}}   & \textbf{0.9772} & 0.9243          & 0.5455          & 0.5995        & 0.9778          & 0.6933          & 0.6608          & 0.8720          & 0.8715          & \textbf{0.9720}   \\ \hline
\multirow{2}{*}{\textbf{3D NR}} & \textbf{SINQ \cite{liu2017binocular}}    & 0.9394          & 0.9362          & 0.8788          & 0.7818      & 0.9429          & 0.8182          & 0.7235          & 0.9030          & 0.8412          & 0.8286      \\
& \textbf{BSVQE \cite{chen2018blind}}   & 0.9265          & 0.9273          & 0.8632          & 0.8781       & 0.9429          & 0.6485          & 0.7471          & 0.9394          & 0.9324          & 0.9429    \\ \hline
\multirow{5}{*}{\textbf{Multi-view FR}} & \textbf{MP-PSNR Full \cite{sandic2015dibr}}  & 0.8750          & 0.8388          & 0.7247    &0.8591      & 0.9404          & 0.8448          & 0.8254  &0.9267   &0.9108     &0.9137    \\
& \textbf{MP-PSNR Reduc \cite{sandic2016multi}} & 0.9083          & 0.8557          & 0.7238     &0.8542     & 0.9404          & 0.8462          & 0.8193  &0.9318   &0.9202     &0.9177     \\
& \textbf{MW-PSNR Full \cite{sandic2015dibr1}}  & 0.8548          & 0.8169          & 0.7238    &0.8464       & 0.9395          & 0.8434          & 0.8322  &0.9347   &0.9260     &0.9467   \\
& \textbf{MW-PSNR Reduc \cite{sandic2015dibr1}} & 0.9020          & 0.8212          & 0.7206    &0.8604        & 0.9511          & 0.8434          & 0.8227  &0.9265   &0.9235     &0.9453  \\
& \textbf{3DSwIM \cite{battisti2015objective}}        & 0.4015          & 0.3842          & 0.5455  &0.7983     & 0.4897          & 0.6063          & 0.5721  &0.7926   &0.6878     &0.6627    \\ \hline
\textbf{Multi-view NR} & \textbf{APT \cite{gu2018model}}   & 0.2182  & 0.0881          & 0.4466          & 0.2855       & 0.3820          & 0.0559          & 0.2918  &0.0381   &0.1303     &0.2207   \\ \hline
\textbf{LFI RR} & \textbf{LF-IQM \cite{8632960}}           & 0.1152          & 0.5222          & 0.6848    &0.6984   &0.6000    &0.1741   &0.1736   &0.3468   &0.3882     &0.7143      \\ \hline
\textbf{LFI NR} & \textbf{Proposed NR-LFQA}     & 0.9483          & \textbf{0.9423} & \textbf{0.8909} & \textbf{0.8903} &0.9429 & \textbf{0.8545} & \textbf{0.8353} & \textbf{0.9515} & \textbf{0.9529} & 0.9429    \\ \hline
\end{tabular}
\end{table*}

In order to demonstrate the effectiveness of our proposed NR-LFQA model, we conduct extensive experiments by using existing 2D, 3D, multi-view, and LFI quality assessment algorithms. In our experiments, we utilize ten 2D-FR metrics \cite{wang2004image,sheikh2006image,zhang2011fsim,wang2003multiscale,wang2011information,chandler2007vsnr,sheikh2005information,damera2000image,mantiuk2011hdr}, three 2D-NR metrics \cite{mittal2012making,mittal2012no,ghadiyaram2017perceptual}, one 3D-FR metric \cite{chen2013full}, two 3D-NR metrics \cite{chen2018blind,liu2017binocular}, five multi-view FR metrics \cite{battisti2015objective,sandic2015dibr1,sandic2015dibr,sandic2016multi}, one multi-view NR metric \cite{gu2018model}, one LFI FR metric \cite{fang2018light}, and one LFI RR metric (i.e. LF-IQM) \cite{8632960}. For all 2D-FR, multi-view FR, LFI FR, LF-IQM, NIQE and APT algorithms, the global predicted score of LFI is obtained by averaging scores of each SAI. We treat the horizontal adjacent SAIs of the LFI as left and right views to test 3D IQA methods. For the rest of the methods, we first extract features on each SAI and then average them. Finally, we predict the LFI quality according to regression methods in their papers. It should be noted that for LF-IQM, we first select the Accurate Depth Map (ADM) \cite{jeon2015accurate} to estimate the corresponding depth maps of original and distorted light field images. Then, we calculate the SSIM value between original and distorted depth maps as objective scores. Finally, the correlation between the objective scores after mapping and subjective ratings can be obtained as the ultimate performance results.

TABLE I shows the performance comparison of objective models on Win5-LID, MPI-LFA and SMART datasets, where bold numbers indicate the best results. We can find that our proposed model outperforms state-of-the-art algorithms. The existing 2D and 3D algorithms only focus on spatial quality prediction, but do not take into account angular consistency degradation. Although multi-view evaluation methods consider angular interpolation distortion, the original intention of the design is to deal with the hole distortion caused by the synthesis, and it is not possible to effectively measure the distortions appearing in the LFI, such as compression distortion. Moreover, spatial texture information is not involved in existing LFI quality evaluation algorithms, and LFI-IQM is greatly affected by depth map estimation algorithms. Therefore, their performance is poor on Win5-LID, MPI-LFA and SMART datasets.
\renewcommand\arraystretch{1.3}
\begin{table}[!htb]
\centering
\caption{Cross Validation Results by Training The Proposed Model on Win5-LID and Testing on MPI-LFA.}
\begin{tabular}{c|ccc}
\hline
\textbf{}    & \textbf{SRCC} & \textbf{LCC}  & \textbf{RMSE} \\ \hline
\textbf{Proposed NR-LFQA} & 0.8389         & 0.7653              & 0.4913        \\ \hline
\end{tabular}
\end{table}
In addition, the Win5-LID and MPI-LFA datasets use different acquisition devices, and the LFIs captured by the TSC in the MPI-LFA are close to that captured by HDCA. This demonstrates that our proposed model is suitable for various scenarios and acquisition devices. Specifically, the proposed method performs well for different content types with variety of SI and CF. Moreover, it also has good performance for rich semantic content.

Apart from conducting experiments on Win5-LID, MPI-LFA and SMART datasets datasets, we also use VALID dataset because it contains both 8bit and 10bit LFIs. As shown in TABLE II, almost all FR algorithms deliver good performance, and our proposed model is superior to all NR algorithms  for all the codecs in VALID dataset. This situation may be caused by that VALID only uses 5 original LFIs, and their distribution is relatively concentrated. As shown in Fig. 3(h), the distribution intervals of CF and SI of VALID are narrower than other datasets.

In order to show the prediction results more clearly, we illustrate the scatter plots of four methods and the proposed model on Win5-LID dataset in Fig. 10. Clearly, the points of NR-LFQA are more centralized than other existing methods and can be well fitted, which demonstrates that the predicted LFI quality scores by NR-LFQA are more consistent with subjective ratings.

To further analyze the performance stability of our proposed model, we visualize the statistical significance. Specifically, we show box plots of the distribution of the SRCC values for each of 1000 experimental trials. Here, we randomly split the entire datasets into training and test sets according to the ratio of 8:2, and test the results of all methods for 1000 times. As shown in Fig. 11, compared with the existing methods, our proposed model results are more concentrated, indicating that NR-LFQA has better stability. The lower the standard deviation with higher median SRCC, the better the performance.

\renewcommand\arraystretch{1.3}
\begin{table*}[!htb]
	
	\centering
	\scriptsize
	\caption{Comparison of The Computation Time against SRCC, LCC, RMSE and OR Performance on Win5-LID Dataset.}
	\begin{tabular}{c|c|c|cccc}
		
		\hline
\textbf{Type}	 & \textbf{Metrics}     & \textbf{Total Computation Time (s)}   & \textbf{SRCC}  & \textbf{LCC}   & \textbf{RMSE} & \textbf{OR} \\ \hline
		
\multirow{10}{*}{\textbf{2D FR}}	 & \textbf{PSNR}     	&0.8188     & 0.6026          & 0.6189          & 0.8031    &0.0045        \\
		
	 & \textbf{SSIM \cite{wang2004image}}      							&2.3068    & 0.7346          & 0.7596          & 0.6650     &0.0000    \\
		
	 & \textbf{MS-SSIM \cite{wang2003multiscale}}    				&3.2937     & 0.8266          & 0.8388          & 0.5566    &0.0000     \\
		
	& \textbf{FSIM \cite{zhang2011fsim}}     							   &14.4056     & 0.8233          & 0.8318          & 0.5675    &0.0045     \\
		
	& \textbf{IWSSIM \cite{wang2011information}}    				 &27.0113    & 0.8352          & 0.8435          & 0.5492   &0.0000    \\
		
	& \textbf{IFC \cite{sheikh2005information}}     					&69.5933      & 0.5028          & 0.5393          & 0.8611  &0.0000   \\
		
	 & \textbf{VIF \cite{sheikh2006image}}      							   &65.9176     & 0.6665          & 0.7032          & 0.7270   &0.0000  \\
		
	& \textbf{NQM \cite{damera2000image}}     						   &16.5390      & 0.6508          & 0.6940          & 0.7362  &0.0045    \\
		
	 & \textbf{VSNR \cite{chandler2007vsnr}}     						 &4.0910    & 0.3961          & 0.5050          & 0.8826  &0.0182  \\
		
	& \textbf{HDR-VDP2 \cite{mantiuk2011hdr}}  						  &115.1300 		&0.5555 &0.6300 &0.7941 &0.0045  \\ \hline
		
\multirow{3}{*}{\textbf{2D NR}}	& \textbf{BRISQUE \cite{mittal2012no}}    &4.4593   & 0.6687          & 0.7510          & 0.5619  &0.0000   \\
		
	& \textbf{NIQE \cite{mittal2012making}}     &8.8498     & 0.2086          & 0.2645          & 0.9861   &0.0045   \\
		
	& \textbf{FRIQUEE \cite{ghadiyaram2017perceptual}}      &2343.0336   & 0.6328          & 0.7213          & 0.5767  &0.0000      \\ \hline
		
\textbf{3D FR}	& \textbf{Chen \cite{chen2013full}}  &1239.3772        & 0.5269          & 0.6070          & 0.8126  &0.0091    \\ \hline
		
\multirow{2}{*}{\textbf{3D NR}}	& \textbf{SINQ \cite{liu2017binocular}}     &309.7299     & 0.8029          & 0.8362          & 0.5124  &0.0000    \\
		
	& \textbf{BSVQE \cite{chen2018blind}}    &396.6745     & 0.8179          & 0.8425          & 0.4801  &0.0000       \\ \hline
		
\multirow{5}{*}{\textbf{Multi-view FR}}	& \textbf{MP-PSNR Full \cite{sandic2015dibr}}  &32.2917		& 0.5335          & 0.4766          & 0.8989   &0.0000       \\
		
	& \textbf{MP-PSNR Reduc \cite{sandic2016multi}} &16.2708	& 0.5374          & 0.4765          & 0.8989   &0.0000      \\
		
	& \textbf{MW-PSNR Full \cite{sandic2015dibr1}} &1.1421	 & 0.5147          & 0.4758          & 0.8993   &0.0000       \\
		
	& \textbf{MW-PSNR Reduc \cite{sandic2015dibr1}} &1.1352		& 0.5326          & 0.4766          & 0.8989   &0.0000     \\
		
	& \textbf{3DSwIM \cite{battisti2015objective}}   &322.9451     & 0.4320          & 0.5262          & 0.8695   &0.0182     \\ \hline
		
\textbf{Multi-view NR}	& \textbf{APT \cite{gu2018model}}     &2626.8449      & 0.3058          & 0.4087          & 0.9332  &0.0045   \\ \hline
		
\textbf{LFI RR}	& \textbf{LF-IQM \cite{8632960}}     &1168.7424      & 0.4503          & 0.4763          & 0.8991  &0.0273   \\ \hline
		
\textbf{LFI NR}	& \textbf{Proposed NR-LFQA}     & 432.1031     & \textbf{0.9032} & \textbf{0.9206} & \textbf{0.3876} & \textbf{0.0000} \\ \hline
	\end{tabular}
\end{table*}

Besides direct comparisons with numerous IQA methods, we also quantitatively evaluate the statistical significance using the t-test \cite{mittal2012no} based on the SRCC values obtained from the 1000 train-test trials for these image quality methods. The null hypothesis is that the mean correlation for row metric is equal to mean correlation for the column metric with a confidence of 95\%. Specifically, a value of `0' indicates that the row and column metrics are statistically indistinguishable (or equivalent), and we cannot reject the null hypothesis at the 95\% confidence level. A value of `1' represents that the row algorithm is statically superior to the column one, whereas a `-1' indicates that the row metric is statistically worse than the column one. From the results on Win5-LID dataset in TABLE III, we also prove that our proposed model is significantly better than all the objective IQA algorithms.

\subsection{Comparison with Other Methods based on EPI}
Since our proposed model uses EPI information, we verify the performance of existing algorithms on EPI for fair comparison. Due to the size property of EPI, some algorithms are not applicable on EPI (e.g. all 3D IQA methods). The FR metrics predict the final score by averaging scores of all EPIs, while the NR metrics extract feature from EPIs and then average these features. TABLE IV shows the results of aforementioned methods on Win5-LID and MPI-LFA datasets, which indicates that our NR-LFQA outperforms existing methods. Additionally, for most algorithms, using EPI information does not yield performance gains. This is because they are not specifically designed to extract information from the EPI.

\subsection{Validity of Individual Quality Component}
We explore the effectiveness of three proposed components ($F_{LCN}$, $F_{GDD}$, $F_{WLBP}$).
The performance values are shown in TABLE V. We can observe that the $F_{WLBP}$ performance is the best on both Win5-LID and MPI-LFA datasets because it can capture subtle changes between SAIs from EPI. This also demonstrates that the weight of $F_{WLBP}$ should be higher in regression.
The reason for the slightly lower performance of $F_{LCN}$ on the MPI-LFA dataset is that only HEVC and GAUSS distortion can significantly affect the spatial quality of the LFI. Moreover, other distortions primarily affect the angular consistency of the LFI.
Since $F_{GDD}$ is insensitive to small angular consistency degradation and MPI-LFA contains many angular distortions with low distortion levels, it has lower performance on MPI-LFA dataset.
Overall, the results demonstrate the validity of our proposed features and the performance is improved after combining features.

\subsection{Robustness Against Distortion Types}
As Win5-LID and MPI-LFA datasets consist of different distortion types, we show how our proposed model performs for each distortion type. The results are listed in TABLE VI. It can be seen that our proposed model outperforms existing objective methods for most distortion types.
Specifically, our proposed model achieves the best performance for typical reconstruction distortions.
The reason is that reconstruction distortions mainly destroy angle consistency and often have little effect on spatial quality. Therefore, existing models are difficult to handle such distortions.

The compression distortion caused by HEVC mainly leads to spatial quality degradation. Therefore, our proposed model is not as good as the FR models. However, our model is still very competitive and has the best performance in the NR methods. Similarly, GAUSS distortion also mainly affects LFI spatial quality. The JPEG distortion in Win5-LID is introduced based on lenslet, and it affects both LFI spatial quality and angular consistency. Our proposed model can simultaneously consider the impact of these two factors, so it is not surprising that our model has the best performance for JPEG distorted LFIs. Overall, the proposed model can achieve the excellent performance against existing objective evaluation schemes involving all the distortion types.

\subsection{Model Generality and Time Complexity}
Since aforementioned methods are generally trained and tested on various splits of a single dataset, it is interesting to know whether the algorithm is dataset dependent. To verify that our proposed model does not depend on the dataset, we choose the same distortion in MPI-LFA and Win5-LID. Specifically, we trained proposed model on Win5-LID dataset, and tested it on the same distortions in MPI-LFA dataset. The results are shown in TABLE VII. Obviously, the proposed model still has good performance, which proves the generality of our proposed model.

As shown in the above table, we compare the proposed NR-LFQA with state-of-the-art quality assessment algorithms regarding to time complexity on Win5-LID dataset. Besides, we also list the SRCC, LCC, RMSE and OR performance values for fair comparison. From TABLE VIII, we can find that the proposed method is verified to have lower computation time compared to the LFI RR metric (i.e. LF-IQM \cite{8632960}). One possible explanation is that different from traditional LFI-QA metrics, our proposed algorithm does not introduce the complex computation of depth map estimation. Generally, our proposed NR-LFQA is in the same level of time complexity compared with state-of-the-art 3D metrics. Moreover, the proposed NR-LFQA demonstrates the best SRCC, LCC, RMSE and OR performance among all schemes.

\section{Conclusion}
In this paper, we have presented a No-Reference Light Field image Quality Assessment (NR-LFQA) scheme. The main contributions of this work are:
1) We propose the first no-reference metric to assess LFI overall quality by measuring the degradation of spatial quality and angular consistency.
2) To measure LFI spatial quality, we analyze the naturalness of light field cyclopean image array.
3) Since the epipolar plane image contains various slopes of lines reflecting angular consistency, we propose gradient direction distribution to measure the degradation of angular consistency and weighted local binary pattern is proposed to capture the subtle angular consistency changes.
4) We compare with state-of-the-art 2D, 3D image, multi-view, and LFI quality assessment methods with our NR-LFQA model. Experimental results show that our model outperforms other metrics and can handle the typical distortions of LFI.

In the future, we will consider to develop a parametric model based on the proposed features and extend our algorithm to light field video quality assessment. Additionally, no-reference light field quality assessment metrics are important for future immersive media processing chain from light field content acquisition, compression, to transmission, reconstruction, and display. We will continue to explore the proposed NR-LFQA for more potential immersive applications, e.g. utilizing NR-LFQA as a loss metric to guide the optimization of synthesizing or compressing light field images.

\bibliographystyle{IEEEtran}
\bibliography{refs}

\end{document}